\documentclass[final,1p,times]{elsarticle}

\usepackage{graphicx}
\usepackage{url}
\usepackage{xurl}
\usepackage{amssymb}
\usepackage{bm}
\usepackage{array}
\usepackage{caption}
\usepackage{subcaption}
\usepackage[hidelinks]{hyperref}

\usepackage{lineno}
\newgeometry{left=2.5cm,right=2.5cm,top=2.5cm,bottom=2.5cm}

\biboptions{comma,square,numbers}

\journal{Science of Remote Sensing}

\begin{document}

\begin{frontmatter}

\title{Satellite Data Shows Resilience of Tigrayan Farmers in Crop Cultivation During Civil War}

\cortext[cor1]{hkerner@asu.edu}

\author[asu]{Hannah R. Kerner\corref{cor1}}
\author[umd]{Catherine Nakalembe}
\author[umd]{Benjamin Yeh}
\author[umd]{Ivan Zvonkov}
\author[umd]{Sergii Skakun}
\author[umd]{Inbal Becker-Reshef}
\author[nasa]{Amy McNally}

\address[asu]{School of Computing and Augmented Intelligence, Arizona State University, Tempe, AZ, United States}
\address[umd]{Department of Geographical Sciences, University of Maryland, College Park, MD, United States}
\address[nasa]{NASA Goddard Space Flight Center, Greenbelt, MD, United States}

\begin{abstract}
The Tigray War was an armed conflict that took place primarily in the Tigray region of northern Ethiopia from November 3, 2020 to November 2, 2022. Given the importance of agriculture in Tigray to livelihoods and food security, determining the impact of the war on cultivated area is critical. However, quantifying this impact was difficult due to restricted movement within and into the region and conflict-driven insecurity and blockages.
Using satellite imagery and statistical area estimation techniques, we assessed changes in crop cultivation area in Tigray before and during the war. Our findings show that cultivated area was largely stable between 2020-2021 despite the widespread impacts of the war. We estimated $1,132,000\pm133,000$ hectares of cultivation in pre-war 2020  compared to $1,217,000 \pm 132,000$ hectares in wartime 2021. Comparing changes inside and outside of a 5 km buffer around conflict events, we found a slightly higher upper confidence limit of cropland loss within the buffer (0-3\%) compared to outside the buffer (0-1\%). Our results support other reports that despite widespread war-related disruptions, Tigrayan farmers were largely able to sustain cultivation. Our study demonstrates the capability of remote sensing combined with machine learning and statistical techniques to provide timely, transparent area estimates for monitoring food security in regions inaccessible due to conflict.

\end{abstract}


\end{frontmatter}

\section{Introduction}
\label{sec:intro}

Armed conflicts profoundly disrupt agriculture and food systems, jeopardizing food security and livelihoods for affected populations \citep{morton2007impact}. Direct impacts include damage to agricultural infrastructure, loss of labor and inputs, market dysfunctions, and prevention of farming activities \citep{flores2005food,nyssen2023crop}. Indirect impacts include environmental degradation, uncontrolled resource exploitation, and displacement of farming communities \citep{suhrke1993pressure}. Quantifying agricultural land use changes is critical for assessing the impact of war on food security in inaccessible conflict regions and targeting humanitarian relief efforts \citep{morales2022co}. 

Satellite remote sensing data can provide qualitative and quantitative insights about the impact of conflict on agricultural land use and production. This remote, real-time data source is especially critical for assessing impacts in regions where traditional ground-based surveys cannot be conducted due to insecurity. 

Most studies using satellite data to assess war impacts on cultivation in other regions have found significant reductions in cultivated land due to war, ranging from approximately 15-30\% across studies. Skakun et al.~\citep{skakun2019satellite} found a 22\% net loss of cropland in regions not occupied by the Ukraine government following the military conflict that began in 2014 in southeastern Ukraine, and that cropland losses were higher (46\%) in areas in the buffer zone surrounding the conflict border. In a study of the more recent war in Ukraine that began in 2022, Ma et al.~\citep{ma2022spatiotemporal} detected concentrations of fallowed croplands in western Kherson and central Luhansk associated with conflict activities. Witmer~\cite{witmer2008detecting} found that the average cultivated area decreased by 25\% in the period after the Bosnian War between 1992-1995. In a study of the armed conflict in Mali, Boudinaud and Orenstein \citep{boudinaud-mali} found that 25\% of areas in Mopti exhibited crop area losses in 2019 compared to pre-conflict years. Olsen et al.~\citep{olsen2021impact} concluded that cropland cultivation decreased by 16\% from 2016 to 2018 as a result of armed conflict in South Sudan. In the first (1994-1996) and second (1999-2009) Chechen wars, Yin et al.~\citep{yin2019agricultural} found higher rates of cropland abandonment in areas closer to conflict events (45\% abandonment within 1 km compared to 6\% greater than 8 km from conflict events) as well as higher rates of re-cultivation in areas farther away from conflict events.
Müller and Kummerle \citep{muller2009causes} and Peterson and Aunap \citep{peterson1998changes} studied the impact of government and economic transitions in Romania after 1989 and Estonia between 1990-1993, respectively. While these transitions were not armed conflicts, they were tumultuous periods with disruptions to the agricultural sector that were associated with 32\% of croplands being abandoned in Estonia and 21\% abandonment in Romania.  

While most studies concluded net losses or negative impacts to agricultural production due to conflict, Li et al.~\citep{li2022civil} found that there was a net \textit{increase} in rural areas of Syria during the civil war. Eklund et al.~\citep{eklund2017conflict} reported a mixture of cropland expansion, abandonment, and other changes in areas of Syria and Iraq occupied by the Islamic State. NASA Harvest \citep{harvest-ukraine} found that despite fears of disruptions to wheat production in Ukraine driven by the Russia-Ukraine war, wheat production in Ukraine in 2023 was similar to pre-war production levels. In a study of cropland abandonment in 14 regions globally, Yin et al.~\citep{yin2020monitoring} found that detecting cropland abandonment was particularly challenging in regions characterized by small field sizes and low-intensity farms such as Nepal and Uganda. 

Together, these studies highlight that the impacts of armed conflict on agricultural land use is highly variable in different contexts. Quantifying changes from satellite data remains challenging, particularly in smallholder farming contexts, and requires a mixture of evidence to shed light on inaccessible regions. 

\begin{figure}
     \centering
     \begin{subfigure}[b]{0.7\textwidth}
         \centering
         \includegraphics[width=\textwidth]{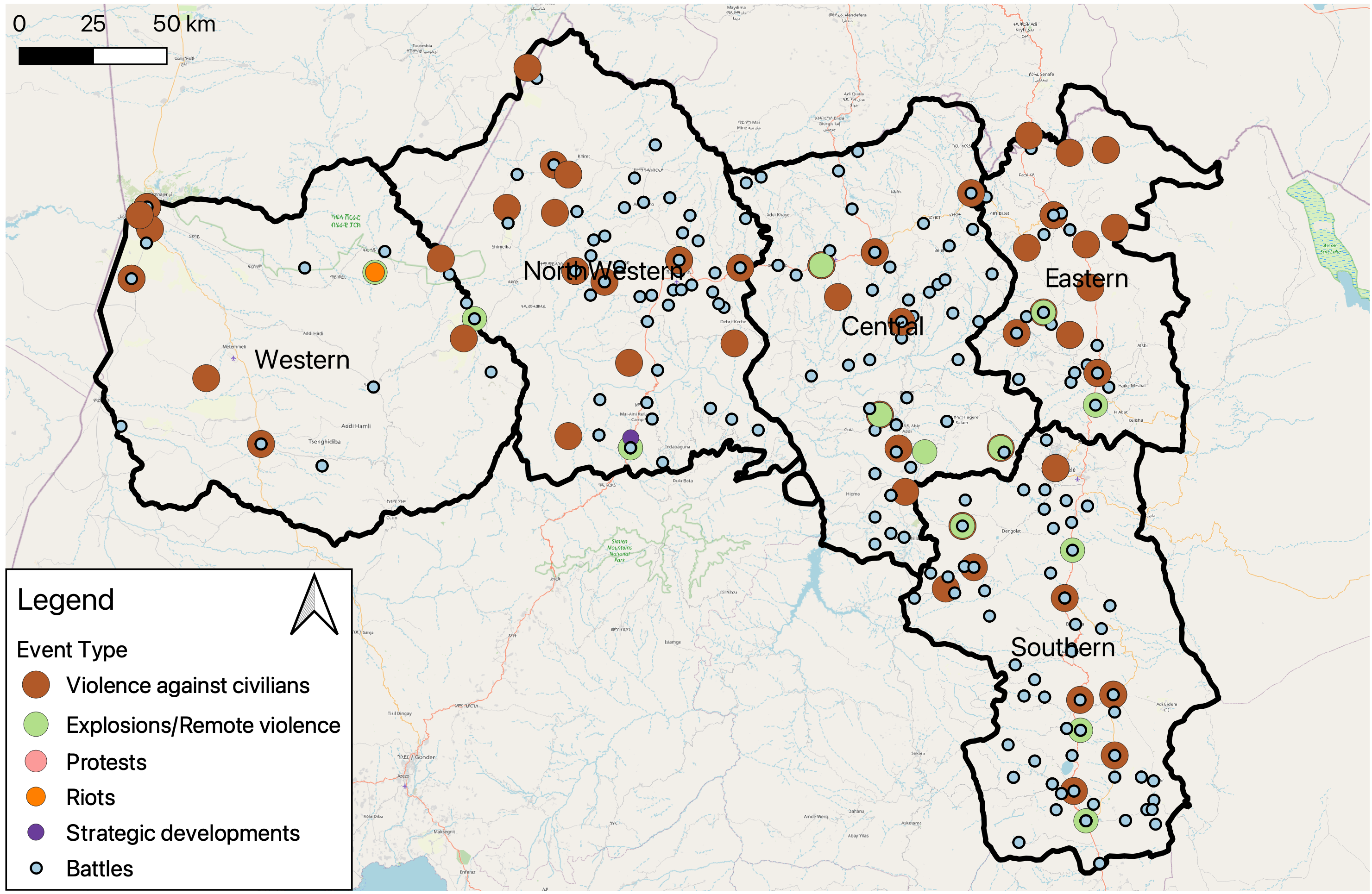}
         \caption{Map of conflict events in ACLED dataset in 2020 and 2021}
         \label{fig:acled-map}
     \end{subfigure} \\
     \begin{subfigure}[b]{0.7\textwidth}
         \centering
         \includegraphics[width=\textwidth]{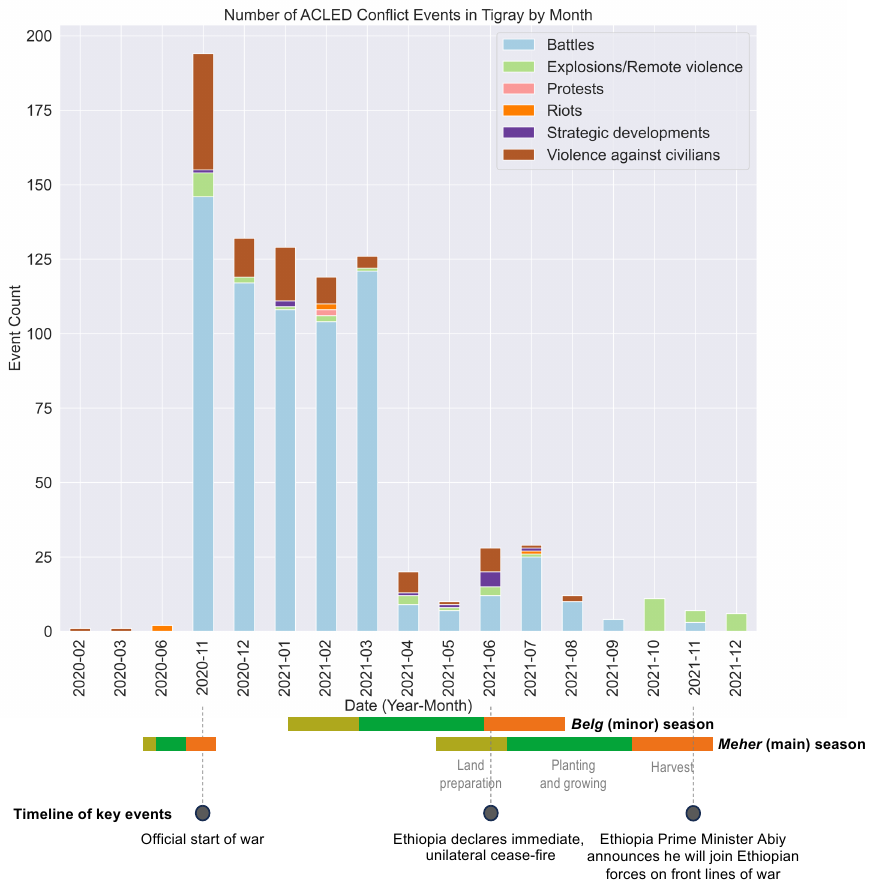}
         \caption{Timeline of events in ACLED database. The crop calendar and timeline of key political events are shown at the bottom, including the official start of the war in November 2020.}
         \label{fig:acled-timeline}
     \end{subfigure}

        \caption{Map and timeline of conflict events from ACLED database}
        \label{fig:timeline}
\end{figure}

The Tigray War was an armed conflict primarily fought in the Tigray region of northern Ethiopia that lasted from November 3, 2020 to November 2, 2022. The conflict involved the Ethiopian federal government and Eritrea on one side and the Tigray People's Liberation Front (TPLF) on the other. Figure \ref{fig:timeline} provides a map and timeline of conflict events recorded by the Armed Conflict Location and Event Data Project (ACLED) in 2020 and 2021 (the timeline considered for our study). Violence and destruction were widespread in the Tigray region, though most conflict events occurred in Southern, North Western, and Central Tigray. Most battles and other incidents occurred between November 2020 and March 2021, and reports of fighting decreased substantially after Ethiopia declared a unilateral cease-fire in June 2021 to allow farmers to till their land and humanitarian aid groups to enter. 

Agriculture is a critical industry in Tigray, providing employment for 80\% of people and constituting 65\% of the land area \citep{un-report}. Disruptions to cultivation could have severe consequences for livelihoods and food security. Thus, quantifying the impact of the war on agriculture is crucial for targeting humanitarian relief to ensure the population's basic needs are met. Many farmers reported delays in ploughing due to insecurity \citep{nyssen2023crop} and  353,000 people in Tigray were reportedly experiencing ``catastrophic'' levels of ``acute food insecurity'' in June 2021 \citep{un-report}. Food insecurity in early 2021 was exacerbated by the loss of food stocks destroyed by armed forces during the start of the war in late 2020 and the blockage of food and humanitarian aid to the region \citep{un-report}. Humanitarian aid was blocked or severely limited from the start of the conflict until April 2021, when aid organizations could deliver food aid for the first time since late 2020. In surveys conducted between April and May 2021, 50-70\% of households in NorthWestern and Central Tigray reported that food assistance was their primary source of food \citep{un-report}. The June 2021 cease-fire gave hope that farmers would be able to cultivate their fields in the 2021 main season, but there were grave concerns and uncertainty about food insecurity and the fate of food production in 2021.

Quantifying the impact of the Tigray civil war on agricultural land use is challenging due to lack of ground-truth data, small field sizes, and heterogeneous landforms. Studies of the impact of the Tigray war on cultivated area found that there was significant and widespread disruption of farming activities particularly in late 2020 and early 2021 including destruction of farm stocks and supplies, delays in ploughing and planting due to threats from soldiers, and market disruptions \citep{weldegebriel2023eyes,nyssen2023crop,un-report}. Nyssen et al.~\citep{nyssen2023crop} reported that despite these disruptions, the effect on cultivated area was relatively moderate because most farmers were able to employ wartime adaptation strategies including shifting the timing and location of farming activities, changing crop types, and leveraging communal aid. Through remote sensing analysis, Weldegebriel et al.~\citep{weldegebriel2023eyes} concluded that there was a modest net loss of 714 km$^2$ of ``well cultivated land'' in highland cropland areas of Tigray (which excludes the Western and parts of the North Western zones) in wartime 2021 compared to pre-war 2019/2020. These studies suggest that the impact of the war on crop cultivation in Tigray was likely moderate. However, there has not yet been a statistically rigorous estimate of the change in cultivated area over the entire region of Tigray before and after the war began.

In this study, we used well-established cropland mapping and area estimation techniques leveraging remote sensing data, machine learning classifiers, and statistical estimators to estimate the change in cropped area in 2020 (pre-war) compared to 2021 (wartime) in Tigray. We used recommended practices for computing scientifically rigorous and transparent estimates of change in land area \citep{olofsson2014good}. While prior work discussed the potential for multifaceted impacts of the Tigray war on food security and the agricultural sector---such as population displacement, changes in crop types or cultivation practices, movement and access limitations, market dysfunction and loss of income, or destruction of food stocks and farming equipment \cite{nyssen2023crop,un-report}---most of these factors are difficult or impossible to determine from satellite observations, especially in the absence of extensive ground-truth data. Our study focused on impacts of the Tigray war on crop cultivation that resulted in a binary change in cultivation status (either cultivated or not cultivated), which could be observed from satellite observations.

Our results provide evidence that complements previous studies of the impact of war on cultivation in Tigray \citep{nyssen2023crop,weldegebriel2023eyes}. Our results show that cultivated area was largely stable between 2020 and 2021, with a relatively small amount of land attributed to cropland gain or loss. Comparing the change in cultivated area in a 5 km buffer of conflict events to the remainder of Tigray, we found a slightly higher upper limit of the confidence interval of the percentage of cropland loss within the conflict buffer (0-3\%) compared to outside (0-1\%). Our findings support the conclusion that Tigrayan farmers were largely able to cultivate crops at a scale comparable to pre-war years despite the widespread impact of the war on farming and other aspects of life.

\section{Methods}
\label{sec:methods}

\subsection{Study area}
The Tigray region spans approximately 50,000 km$^2$ in northern Ethiopia with an estimated population of approximately 5 million as of 2019 \citep{annys2021tigray, CSAEthiopia2017}. Tigray shares a border with Eritrea to the north and Sudan to the west. Tigray contains five zones: Eastern, North Western, Western, Central, and Southern. The terrain in Tigray is very mountainous and hilly, with extensive terracing used for agriculture. Elevations range from 500 meters in the northeast to almost 4,000 meters in the southwest. The climate is generally semi-arid, with average annual rainfall between 400-800 mm concentrated in the summer months (June to September). An estimated 80\% of the population works in agriculture and around 65\% of the land area is devoted to cultivation \citep{annys2021tigray}. The Central and Western zones have more intensive agriculture due to higher rainfall and fertile soils. The majority of crop cultivation in Tigray is rainfed with two growing seasons: \textit{meher} (June-September), which is the main growing season, and \textit{belg} (February-May), which is the minor growing season. The majority of agricultural land in Tigray is used to grow teff, barley, wheat, maize, sorghum, finger millet, oats, and sesame \cite{Ethiopia2015AgriculturalSurvey}. For most crops, planting occurs between May to July, vegetative growth between July and September, and harvesting between September and December \citep{annys2021tigray}.

\subsection{Conflict data}
\label{sec:conflict-data}
We obtained a geographically and temporally referenced dataset of conflict events that occurred in Tigray in 2020 and 2021 from the Armed Conflict Location and Event Data Project (ACLED). ACLED compiles data from various sources including media reports, NGOs and international organizations, government reports and statements, social media and online platforms, academic literature, eyewitness accounts, and local sources. We included all event types provided in the ACLED dataset except ``Peaceful Protests.'' This resulted in a dataset of 831 events with multiple event types: 666 battles, 104 violence against civilians, 43 explosions/remote violence, 11 strategic developments, 5 riots, and 2 protests.

Figure \ref{fig:timeline} shows a map and timeline of these events. From Figure \ref{fig:acled-timeline}, it can be seen that the majority of conflict events occurred prior to the start of the 2021 main growing season. This plot also shows that Ethiopia's declaration of an immediate, unilateral cease-fire occurred during the land preparation phase. However, some battles and other events were still reported after the cease-fire. Of the five zones in Tigray, the majority of the conflict events in the 2020-2021 time period were recorded in the Southern zone (247), followed by the North Western (199), Central (196), Eastern (117), and Western (72) zones. 

November 2020 marked the official start of the war and was the month with the most recorded conflict events during the study period. This also coincided with the harvest period of the main meher season of 2020. Despite this outbreak, final crop yields estimated at the regional and national level were reportedly favorable, with the exception of localized conflict-affected areas which were reported to have lost an estimated 90\% of the main 2020 season harvest due to pillaging and burning by armed forces \cite{FEWSNET2021EthiopiaAlert}. Based on the conflict timeline and previous reports (e.g., \cite{nyssen2023crop,FEWSNET2021EthiopiaAlert}), we expected that the conflict might have some impact on the minor belg 2021 season but were uncertain about the likelihood of impact on the major meher 2021 season. In this study, we did not differentiate between estimated area of crops cultivated during the belg versus meher season since our goal was to estimate the change in the total cultivated between 2020 and 2021. Furthermore, estimating changes between the individual seasons would be difficult since farmers may plant the same fields in different seasons each year.

\begin{figure}
    \centering
    \begin{subfigure}[b]{0.9\textwidth}
         \centering
         \includegraphics[width=\textwidth]{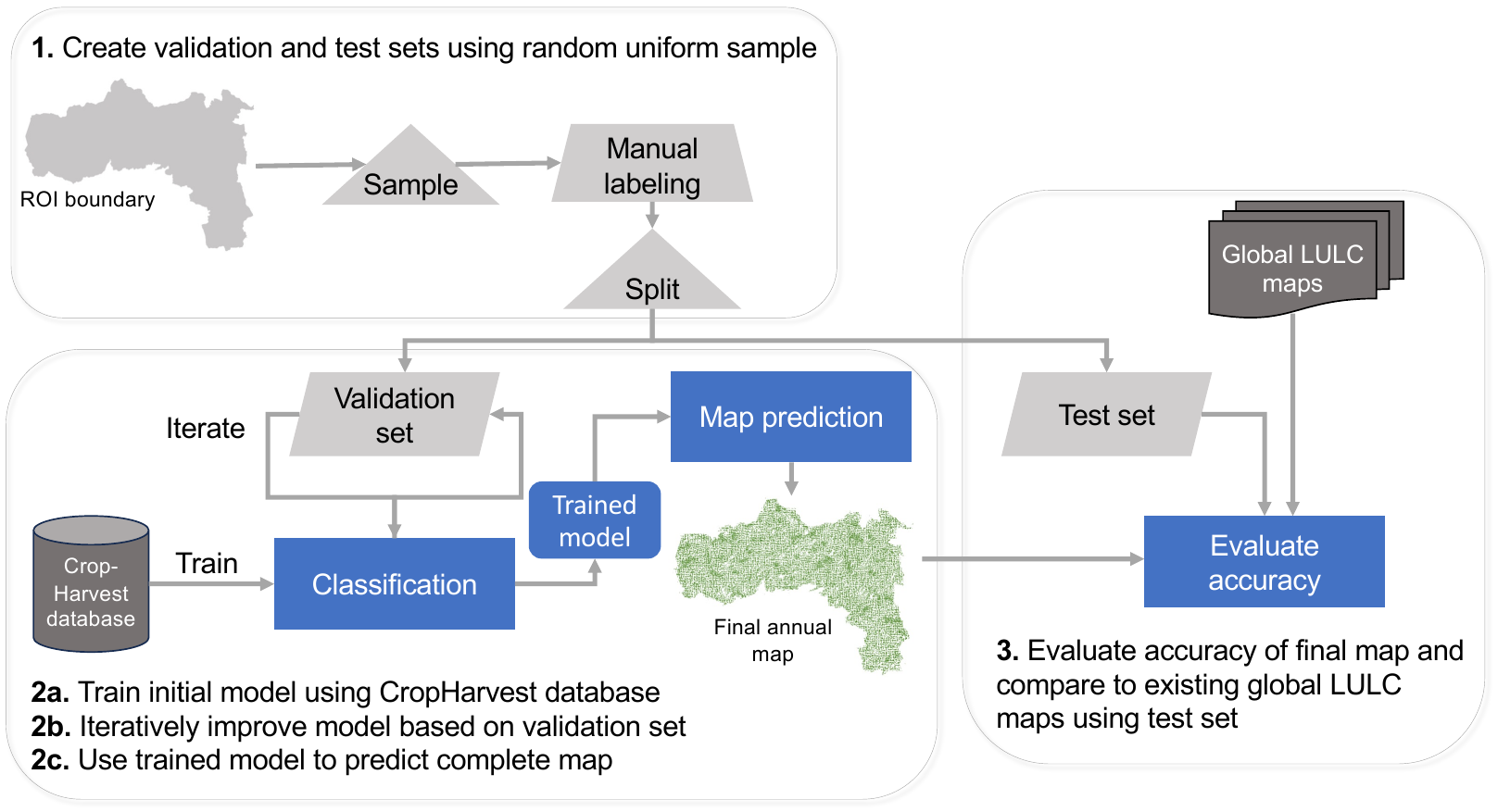}
         \caption{Process diagram for creating annual cropland maps}
         \label{fig:mapping-workflow}
     \end{subfigure}
     \begin{subfigure}[b]{0.9\textwidth}
         \centering
         \includegraphics[width=\textwidth]{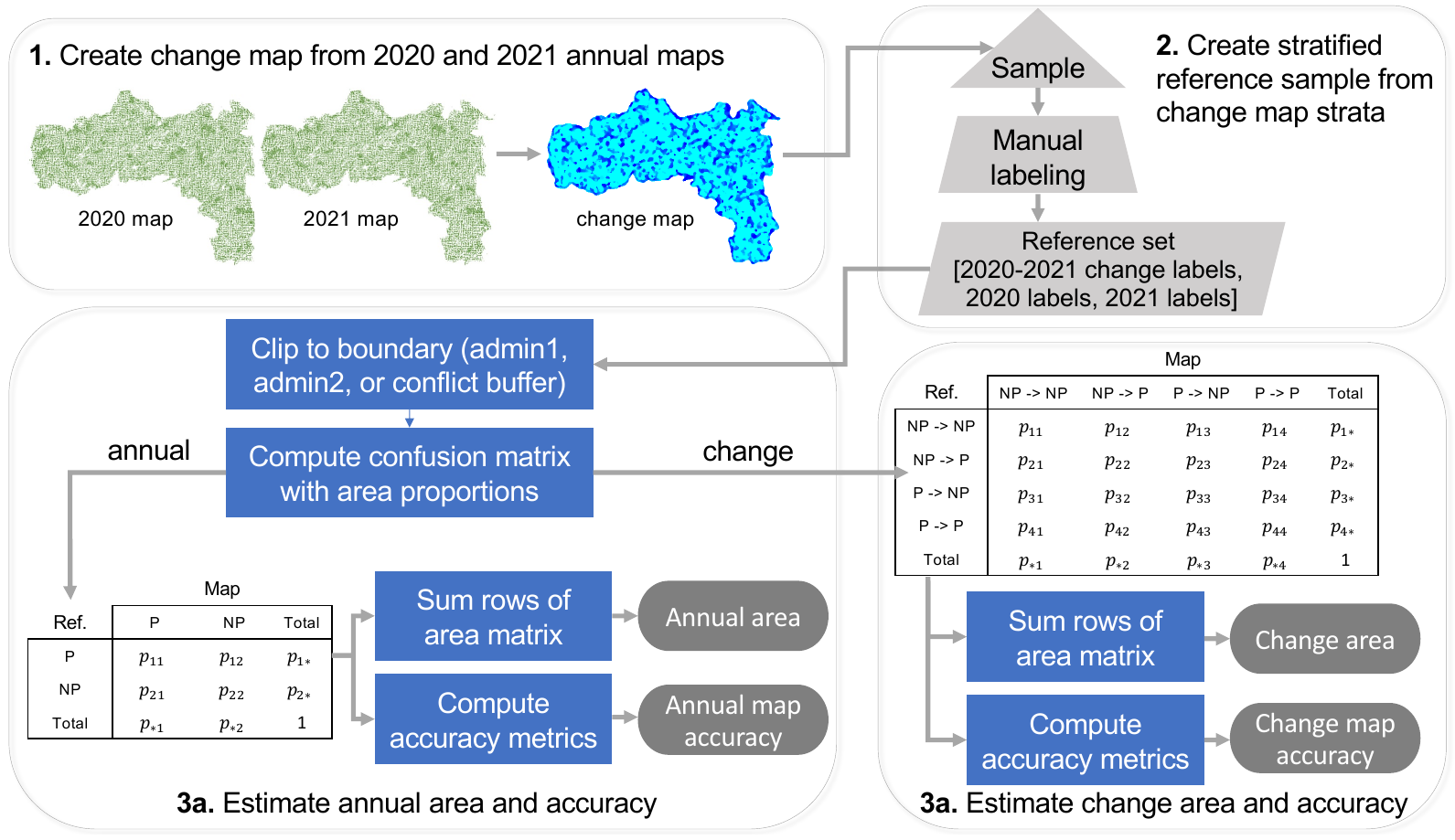}
         \caption{Process diagram for estimating annual cropland area, change area, and map accuracy}
         \label{fig:area-workflow}
     \end{subfigure}
     \caption{Process diagrams for creating annual cropland maps and computing area estimates. Map images are for illustration purposes and do not represent actual results. }
     \label{fig:workflow}
\end{figure}

\subsection{Cropland mapping} %
\label{subsec:cropland-mapping}
\subsubsection{Classification model} 
Figure \ref{fig:mapping-workflow} summarizes the workflow we used to create annual cropland maps for 2020 and 2021.
We used the long short term memory (LSTM) recurrent neural network architecture from Kerner et al.~\citep{rapidresponse} to create a binary cropland map for 2020 and 2021 separately. The LSTM architecture has one base hidden layer, which is followed by two classification heads: one two-layer head is used to classify training instances within Tigray, and one single-layer head is used to classify global training instances (outside of Tigray). We used a batch size of 128 instances. We used early stopping with a patience of 10 epochs and proceeded with model weights from the training checkpoint with the highest F1 score evaluated on the validation set. Training concluded after 16 epochs for the 2020 model and 13 epochs for the 2021 model (one epoch represents one complete pass through the entire training dataset). We used the Adam optimizer with default parameters to tune weights during training with learning rate set to $0.001$. We set the $\alpha$ hyperparameter that balances the global and local samples in the loss function to $10$ \citep{rapidresponse}. We used dropout during training with a dropout probability of $0.2$. We used the trained model for each year to predict the full cropland map for Tigray in 2020 and 2021 using the OpenMapFlow python library \citep{openmapflow}. Our code used to create the cropland maps in this study can be found at \url{https://github.com/nasaharvest/crop-mask}.

\subsubsection{Cropland change mapping}
We created a cropland \textit{change} map for 2020-2021 by combining the cropland maps created for 2020 and 2021 separately. This resulted in a map with four change classes: stable non-crop, stable crop, crop gain, crop loss. Using the annual cropland maps to create a change map, rather than training a model to classify four change classes directly, allowed us to leverage a large training dataset available for binary cropland classification (see Section \ref{subsec:model-datasets}). Since large labeled datasets for four-class cropland change are not available, direct change classification would have required us to label a large dataset of training samples for the change classes, which would have been extremely time-consuming.

\subsubsection{Satellite data}
We used the pixel-level time series input data format from the CropHarvest database \citep{cropharvest} for the LSTM classifier. The 12-month time series consists of aggregated observations in a given pixel over each month from February 1 to January 31 for the year of interest (2020 or 2021) to allow the model to observe the full growing cycle in either the belg or meher season. We included observations from Sentinel-2, Sentinel-1, and the Shuttle Radar Topographic Mission (SRTM) Digital Elevation Model (DEM). Note that while CropHarvest includes meteorological data (precipitation and temperature) from the ERA5 dataset, we omitted this dataset because we found that it introduced artifacts in the predicted map.
We used all bands from Sentinel-2 except B1 (coastal aerosol) and B10 (cirrus) and added NDVI (computed from B8 and B4). This resulted in 12 channels from Sentinel-2. We aggregated the 5-day observations to monthly by taking the least-cloudy pixel per 30-day window using the method described by Schmitt et al.~(2019) \cite{cloudfree}. We used top-of-atmosphere observations (Level 1A) since this was required by the cloud-filtering algorithm.  We used the VV and VH channels from the Sentinel-1 synthetic aperture radar dataset which we aggregated to monthly by taking the median of observations over each month. We used the SRTM DEM to compute elevation and slope features. In total, our input data had 16 channels for each of 12 timesteps for a given pixel (note that elevation and slope are constant over all timesteps). All datasets were pre-processed, resampled to 10 m/pixel, and exported to Google Cloud Storage using Google Earth Engine. The preprocessing code for each dataset can be found at \url{https://github.com/nasaharvest/openmapflow/tree/main/openmapflow/eo}.

\subsubsection{Model training, validation, and test data}
\label{subsec:model-datasets}
Our training dataset consisted of approximately 100,000 total instances with binary labels of crop or non-crop (about 5,000 of which were in Tigray and thus used to train the local classification head). The training dataset was constructed primarily from publicly accessible data in the CropHarvest dataset (\url{https://zenodo.org/doi/10.5281/zenodo.5021761}). Since our training data contained a mix of global and local (within Tigray) samples, we do not expect our model to be biased toward detecting cropland in a specific season.

The validation and test datasets can be found at \url{https://data.harvestportal.org/dataset/annual-and-change-maps-for-tigray-2020-2021}. We created validation and test datasets for choosing hyperparameter settings and evaluating final performance metrics (respectively) using a random uniform sample of points in Tigray. Since points were randomly sampled within Tigray, we expect the proportion of cropland in this sample to represent the approximate proportion of cropland area in the region, including the approximate proportion of planting in the meher and belg seasons typically observed in the region. We created independent samples for each year. Each point was labeled by the authors and other trained individuals. Annotators answered the question ``Does this point contain active cropland?'' for each point. We describe our labeling procedure used to answer this question in Section \ref{subsec:labeling}. Each point was labeled by at least two different annotators and we retained only the points for which both (or a majority of) labelers agreed. This resulted in a validation set of 440 labels (140 crop and 300 non-crop) for 2020 and 289 labels (93 crop and 196 non-crop) for 2021, and a test set of 425 labels (127 crop and 298 non-crop) for 2020 and 297 labels (109 crop and 188 non-crop) for 2021.

\subsubsection{Post-processing}
The University of Ghent and Mekelle University published a small ground-truth survey dataset of 161 fields surveyed in Tigray at the end of August 2021 \citep{ghentdata2021}. The team of geographers visited six districts in Eastern and Southern Tigray (including Mekelle), spanning ecoregions with diverse biophysical and agro-ecological characteristics. Fields surveyed were typically near (hundreds of meters away from) main roads. Selection criteria for fields included rainfed farmland (no irrigation); no buildings, iron sheets, or trees on the land; and field size at least $30$ m $\times$ $30$ m, and no intercropping. Surveyors recorded point locations in the center of each field. Data for each field was recorded by observing and talking to farmers who were present on the land. 

Surveyors recorded the crop type being grown in each field as well as the crop condition (e.g., good, medium, poor) and notes about the nearby farmlands. 
Of the 34 fields recorded as fallow, 20 were recorded as ``fallow with weeds or grass'' and 14 as ``fallow (no vegetation at all)'' (meaning the field had been plowed but not seeded). The 127 remaining cultivated fields included teff (33), maize (7), sorghum (5), other cereals such as wheat, barley, or millet (62), oil crops such as flax or niger seed (12), legumes such as beans or peas (6), and potato (2).

\begin{figure}
    \centering
    \includegraphics[width=\textwidth]{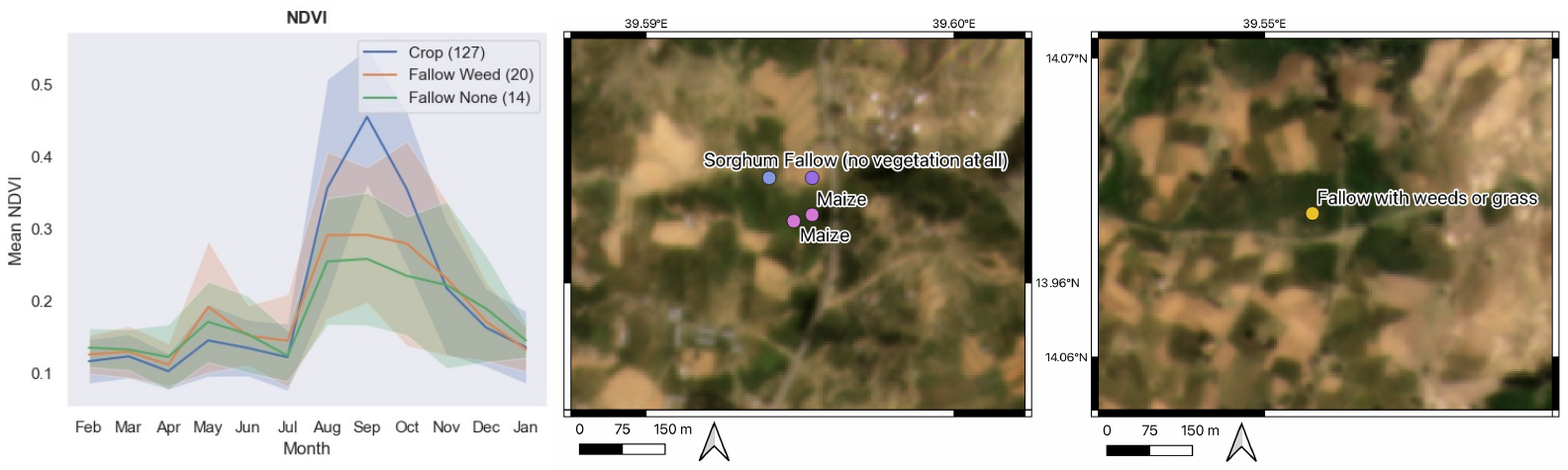}
    \caption{Left: 12-month time series of Sentinel-2 NDVI for ground-truth plots annotated as cultivated (``Crop''), fallow with weeds (``Fallow Weed''), or fallow with no vegetation at all (``Fallow None''). Solid lines depict the mean and shaded regions depict the standard deviation of the time series for all samples. Right: Visualizations of example ground-truth plots in PlanetScope September 2021 basemap. The plot labeled ``Fallow with weeds or grass'' has a similar appearance to fields labeled with crop types, while the field labeled ``Fallow (no vegetation at all)'' is easier to distinguish. Note that our map classification used Sentinel-2 inputs (left-most time series) while our response design used PlanetScope data for photointerpretation of reference sample labels (similar to those shown on the right).}
    \label{fig:ghent-data}
\end{figure}

To better understand the difficulty of distinguishing cultivated from fallow fields (which may resemble cultivated fields due to the presence of weeds or other vegetation) in satellite observations, we plotted the 12-month time series profiles for these ground-truth points in each of the 12 Sentinel-2 channels that were used for our model input. Figure \ref{fig:ghent-data} shows the NDVI time series and \ref{fig:ghent-data-all} shows the time series for all Sentinel-2 bands. Figure \ref{fig:ghent-data-all} shows that the time series for all bands is visually very similar for the cultivated and fallow fields surveyed. This observation, in addition to visual inspection of predicted crop maps in Tigray, suggested that our predicted crop maps may overestimate cultivated area in Tigray by classifying some fallow fields as cultivated.

In Figure \ref{fig:ghent-data}, some differentiation between cultivated and fallow fields can be seen in the NDVI time series: the peak of the NDVI time series appears substantially higher on average than the peak reached in fallow fields. This also makes intuitive sense: we would expect fallow fields to have lower biomass than cultivated fields, since the pattern of vegetation in fallow fields should be sparse and unmaintained. Thus, we hypothesized that for a given predicted cropland map, within the pixels predicted as cropland there will be a minority sub-population of pixels that reached lower peak NDVI than the majority of the pixels which may correspond to fallow fields (or other erroneous predictions). To identify and remove these potential false positives, we computed the mean and standard deviation of the peak NDVI value for all positive (cropland) predictions in a given map. We then identified which of those pixels had NDVI below $n$ standard deviations from the mean (i.e., $\mu - n\sigma$). We then re-classified those ``anomalous'' positive pixels to the non-crop class.

\begin{figure}
    \centering
    \includegraphics[width=\textwidth]{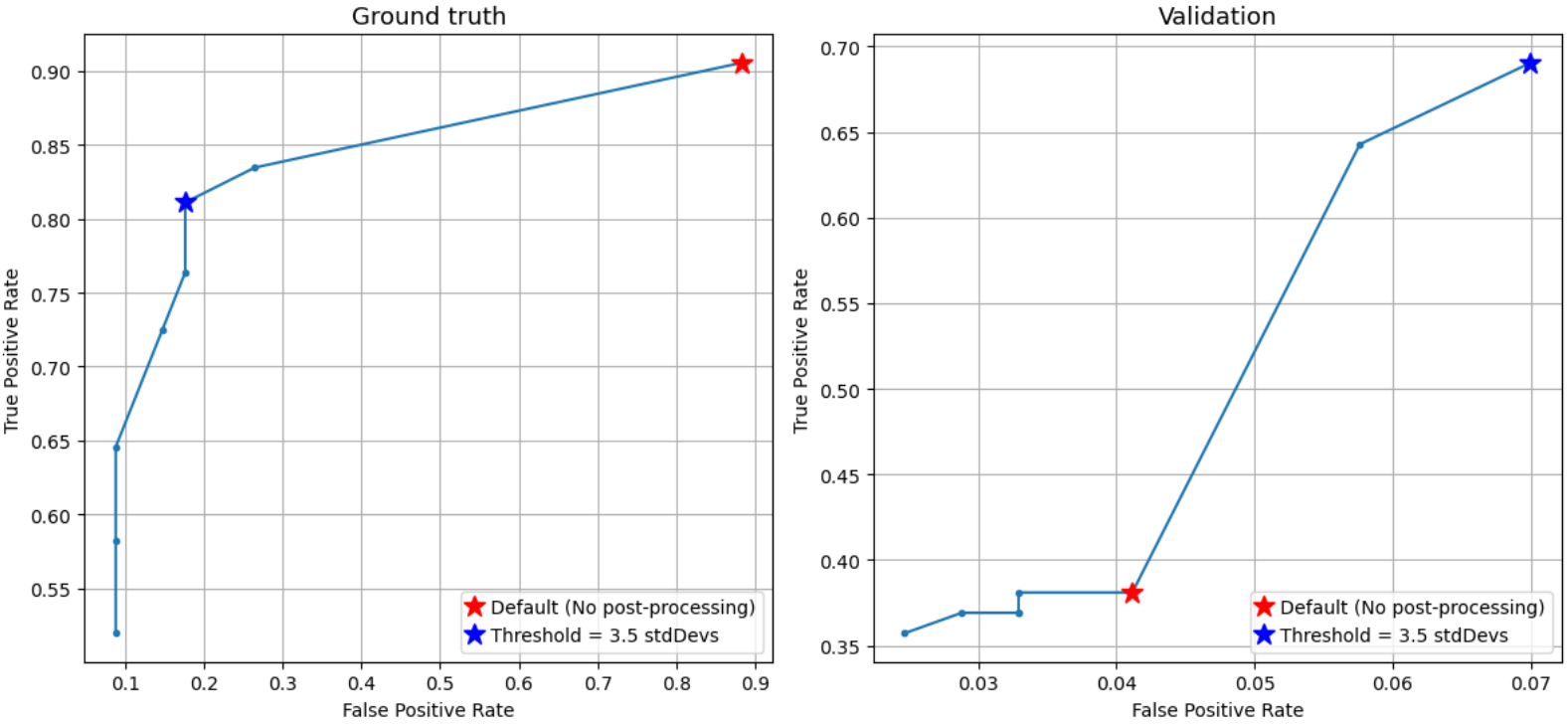}
    \caption{True positive rate (TPR) versus false positive rate (FPR) for range of thresholds used for post-processing algorithm, evaluated for the ground-truth dataset and validation dataset for Tigray in 2021. %
    }
    \label{fig:postprocessing-threshold}
\end{figure}

We evaluated the effect of using $n \in \{0, 0.5, 1.0, 1.5, 2.0, 2.5, 3.0, 3.5, 4.0\}$ for this post-processing algorithm on the true positive rate (TPR) versus false positive rate (FPR) for the ground-truth dataset and validation dataset for Tigray in 2021. Figure \ref{fig:postprocessing-threshold} shows a plot of these values. We found that a threshold of $n=3.5$ offered the optimal tradeoff between TPR and FPR for the ground-truth dataset. This post-processing threshold also increased the TPR on the validation dataset without significantly increasing the FPR. Thus for our final cropland maps in 2020 and 2021, we applied this post-processing algorithm with $n=3.5$ to reduce false positive predictions. 

\subsection{Comparison to existing land cover maps}
\label{subsec:comparison}
Some prior studies used existing global land cover maps to derive a cropland map for change assessments instead of creating a custom classification map as in our study. For example, Weldegebriel et al.~\citep{weldegebriel2023eyes} used the WorldCover 10 m/pixel global land cover map for 2020 \citep{zanaga2021} to assess changes in cultivated area between 2020-2021 in Tigray.

To assess the accuracy of our custom map classification compared to existing land cover maps, we evaluated multiple global land cover maps using our independent test set following the methodology described in Kerner et al.~\citep{kerner2023accurate}. We only included maps that were created in 2017 or later to ensure temporal coherence: WorldCover \citep{zanaga2022esa,worldcover-data}, Digital Earth Africa \citep{burton2022co,dea-data}, Copernicus \citep{buchhorn2020copernicus,cgls-data}, Dynamic World \citep{brown2022dynamic}, ASAP \citep{rembold2019asap,asap-data}, GLAD \citep{potapov2022global,glad-data}, and Esri LULC \citep{karra2021global,esri-data}. If multiple years were provided by a land cover product, we used the closest year to our reference data year.

\subsection{Area estimation}
\label{sec:area-estimation}
Reliable area estimates cannot be obtained by pixel counting (i.e., summing the pixels in each class and multiplying by the area covered by each pixel) from maps that inevitably contain classification errors. Sample-based area estimates are considered good practice for producing unbiased estimates of land area \citep{olofsson2014good}. Mapped areas obtained from pixel counting and sample-based areas may differ significantly \citep{olofsson2020mitigating}. %
We used the best practices outlined in Olofsson et al.~\citep{olofsson2014good} to compute unbiased area estimates using a reference sample stratified by the classification map classes. Figure \ref{fig:area-workflow} summarizes the workflow we used to estimate annual cropland area and change in cropland area between two years.

\subsubsection{Sampling and response design}
\label{subsec:sampling-design}
We created a stratified random sample of $10 \times 10$-meter pixels using the four strata in the 2020-2021 cropland change map classes. We drew 799 samples total (sampling rate 0.00015\% in Tigray) using the following allocation to the four strata: 131 for ``stable planted,'' 404 for ``stable not-planted,'' 134 for ``planted gain,'' and 130 for ``planted loss.'' We pre-allocated 100 samples to each change stratum, then allocated the remaining samples approximately proportionally to the map area for each stratum. Pre-allocation helps reduce the uncertainties in area estimates for change classes resulting from low sample sizes \citep{olofsson2014good}. 

Each point was labeled by the authors and other trained individuals. Annotators answered two questions for each point: ``Does this point contain active cropland in 2020?'' and ``Does this point contain active cropland in 2021?'' Each reference sample was independently labeled by at least two annotators. For samples that did not have a unanimous label agreement for both questions, we convened expert annotators as a group to discuss and decide the label to assign for each year. We describe our labeling procedure used to answer this question in detail in Section \ref{subsec:labeling}.

\subsubsection{Satellite interpretation of reference points}
\label{subsec:labeling}
The labels assigned to the reference samples described in Sections \ref{subsec:model-datasets} and \ref{subsec:sampling-design} have a significant impact on the area estimates and performance metrics derived from those samples. In this section, we detail our procedure for determining whether a given reference point for a given year should be labeled as containing active cropland or not. We provide several examples to illustrate how we determined these labels in \ref{sec:reference-examples}.

We used Collect Earth Online (CEO) as our primary labeling interface. We created a CEO project for each set of reference points and configured the project's Imagery Options to include the 3-5 m/pixel Planet Monthly Mosaics, 10-20 m/pixel Sentinel-2 monthly composite (displaying the Agriculture band combination -- a false-color composite of the SWIR-1 (B11), near-infrared (B8), and blue (B2) bands -- by default), and the 0.5-1 m/pixel Mapbox Satellite layers. The Planet and Sentinel-2 layers allow users to view the imagery for a specific month/year, while the Mapbox layer is a basemap and the date cannot be configured. Thus, we instructed labelers to base their interpretation primarily on Planet and Sentinel-2 since the date of the Mapbox images was unknown -- however, the high-resolution Mapbox basemap can still provide additional context for the coarser Planet and Sentinel-2 layers. 

For points where additional context might be helpful, labelers were instructed to click ``Download Plot KML'' in the CEO interface and load the point in Google Earth Pro on their desktop computer. In Google Earth Pro, labelers used the ``Historical Imagery'' tool to view any high-resolution imagery (typically provided by Maxar or Airbus/CNES) available for the point. While high-resolution images during the growing season of the reference year were rarely available, high-resolution images acquired at any point (or multiple points) in time can provide additional helpful context for interpreting the coarser imagery in CEO (similar to the Mapbox basemap). In Google Earth Pro, labelers can also change the viewing angle of the point and use the terrain layer to better visualize the landscape in 3D -- this is particularly helpful for points on steep slopes or ridges unlikely to host agriculture.

The majority of agricultural land in Tigray is used to grow teff, barley, wheat, maize, sorghum, finger millet, oats, and sesame \cite{Ethiopia2015AgriculturalSurvey}. These are row crops that are planted and harvested on an annual cycle, necessitating yearly tilling, sowing, and management. Thus, as in Skakun et al.~\citep{skakun2019satellite}, we defined cropland as ``a piece of land that is sown/planted and harvestable at least once within the 12 months after the sowing/planting date.'' This is also consistent with the Ethiopia Central Statistics Agency's definition of annual/temporary crops as ``crops, which are grown in less than a year’s time, sometimes only a few months with an objective to sow or replant again for additional production following the current harvest. Continuously grown crops planted in rotation are also considered as temporary crops since each is harvested and destroyed by ploughing in preparation for each successive crop'' \cite{Ethiopia2015AgriculturalSurvey}. Our definition does not include permanent (perennial) crops, since these represent a minority of production in Tigray \cite{Ethiopia2015AgriculturalSurvey}. Our definition also does not include livestock or pasture.

Labelers typically stepped through the Planet Monthly Mosaic for each month spanning the crop calendar for Tigray (considering both the belg and meher seasons), zooming in/out and panning around to assess the point locally and compared to nearby fields or landscapes. They did the same for the Sentinel-2 layer and compared the Mapbox layer to the Planet and Sentinel-2 layers for context. If needed, they downloaded the point's KML file and viewed historical imagery available in Google Earth Pro. Labelers looked for the following indicators of active cultivation of annual crops:
\begin{itemize}
\itemsep 0em
    \item Approximately rectilinear shapes of fields
    \item Relatively uniform color and texture within field boundary suggesting management
    \item Dark brown color during land preparation/planting months indicating plowing
    \item Changes in color and texture aligned with crop calendar: e.g., increase in greenness during months between sowing and vegetated phases, dark green color during peak vegetation month(s), or rapid changes from green to brown in harvest months
    \item Presence of harvest piles at the end of the growing season \cite{xu2024harvestnet}
\end{itemize}

In contrast, fallow (inactive) fields tend to be characterized by:
\begin{itemize}
\itemsep 0em
    \item Lack of uniformity in color and texture
    \item Sparse, mottled, or un-managed appearance
    \item Brown, beige, or a lighter green color indicating the presence of bare soil, weeds, or residual plants
    \item Lack of significant change in color and texture throughout the growing season
\end{itemize}

While all of these characteristics may not be visible in the imagery for every point, labelers looked for a combination of these characteristics to decide the label for a point. We illustrate some of these characteristics in the examples provided in \ref{sec:reference-examples}.

\subsubsection{Cultivated area in Tigray region and zones}
We used the reference samples to compute confusion matrices between the map class and reference labels, which were then used to compute map accuracy, area estimates and uncertainties, and investigate changes between 2020-2021 that could be seen visually in satellite images \citep{olofsson2014good}. We computed area estimates for the four change classes representing transitions from 2020 to 2021 as well as annual cropland area estimates for 2020 and 2021 separately. For the annual area estimates, we used the reference labels from each year provided by the two-year change reference samples. We computed the annual and change area estimates for the Tigray region as well as within each of the five zones in Tigray. For each zone, we clipped the classification map to the zone boundary and filtered the reference samples to use only the samples within a given zone to estimate area. The code used to compute annual area estimates is available at \url{https://github.com/nasaharvest/crop-mask/blob/165d01cc3114427fac59f28ec2eeee5c97de25fc/notebooks/crop_area_estimation.ipynb}. The code used to compute change area estimates is available at \url{https://github.com/nasaharvest/crop-mask/blob/359708379fb7f064e1248ff35daaf56e27eee97a/notebooks/ethiopia_tigray_change_area_estimation.ipynb}. The reference samples can be accessed via Github following the paths recorded in the change area estimation notebook.

\subsubsection{Cultivated area in proximity to conflict events}
To evaluate whether cropland change was concentrated in locations where conflict events directly occurred, we computed the four-class change area estimates inside and outside of a buffer surrounding locations of reported conflict events. Using the ACLED conflict event database described in Section \ref{sec:conflict-data}, we used QGIS to create a 5 km buffer surrounding each event location and then dissolved the boundaries to create a single geometry. We chose 5 km as the buffer size based on previous studies that used similar buffer sizes to assess conflict impacts on agriculture \citep{olsen2021impact,skakun2019satellite,weldegebriel2023eyes}. We estimated cropland change inside this boundary by clipping the change map and filtering the reference samples to the buffer geometry. To estimate cropland change outside of the buffer zone, we created a new geometry representing the difference between the Tigray region boundary and the buffer zone boundary. We then estimated cropland change in the outside-buffer boundary by clipping the change map and filtering the reference samples to the geometry.

\section{Results}
\label{sec:results}

\subsection{Map classification and accuracy assessment}
\begin{figure}
    \centering
     \begin{subfigure}[b]{0.49\textwidth}
         \centering
         \includegraphics[width=\textwidth]{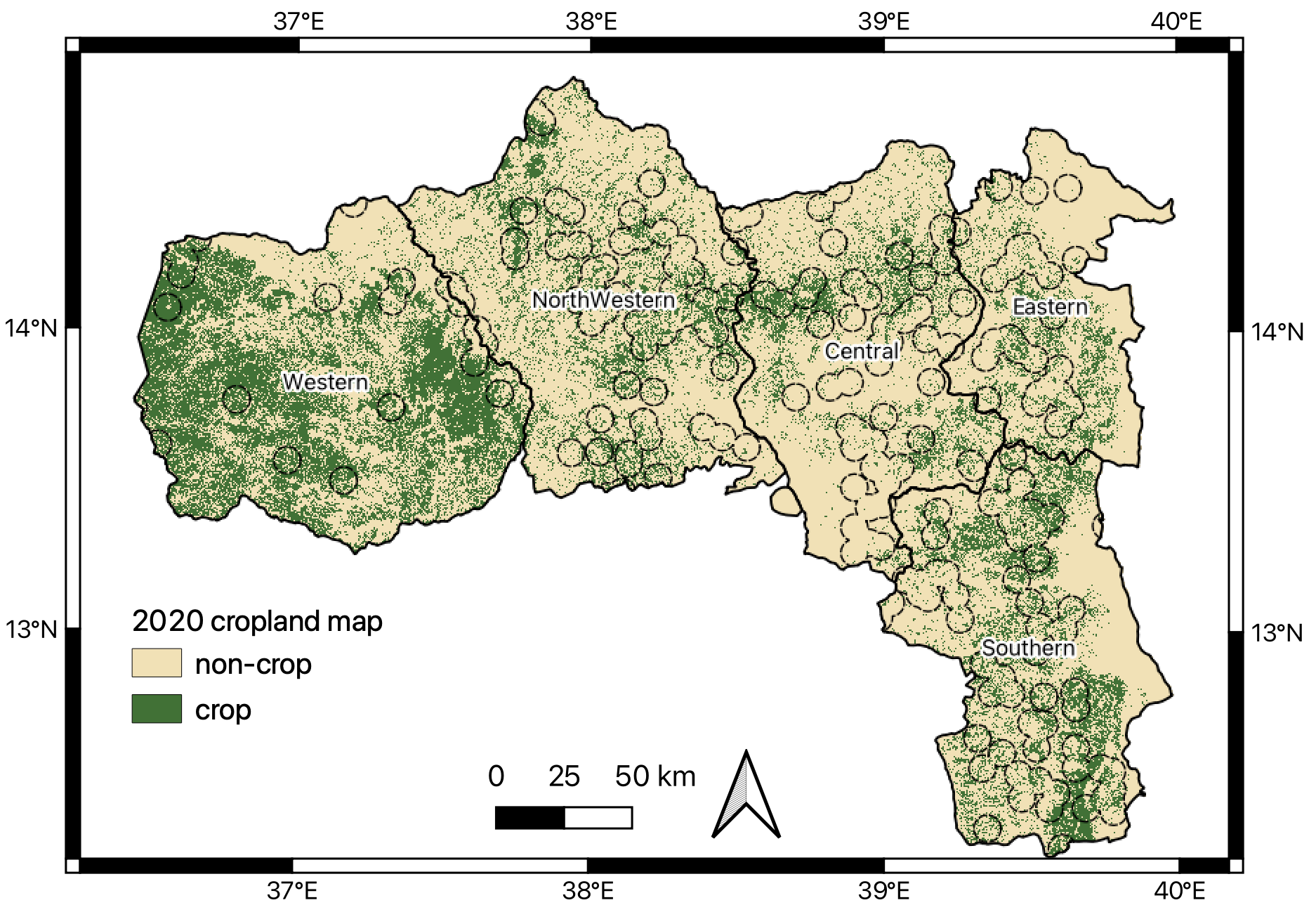}
         \caption{2020 cropland map}
         \label{fig:map2020}
     \end{subfigure}
     \begin{subfigure}[b]{0.49\textwidth}
         \centering
         \includegraphics[width=\textwidth]{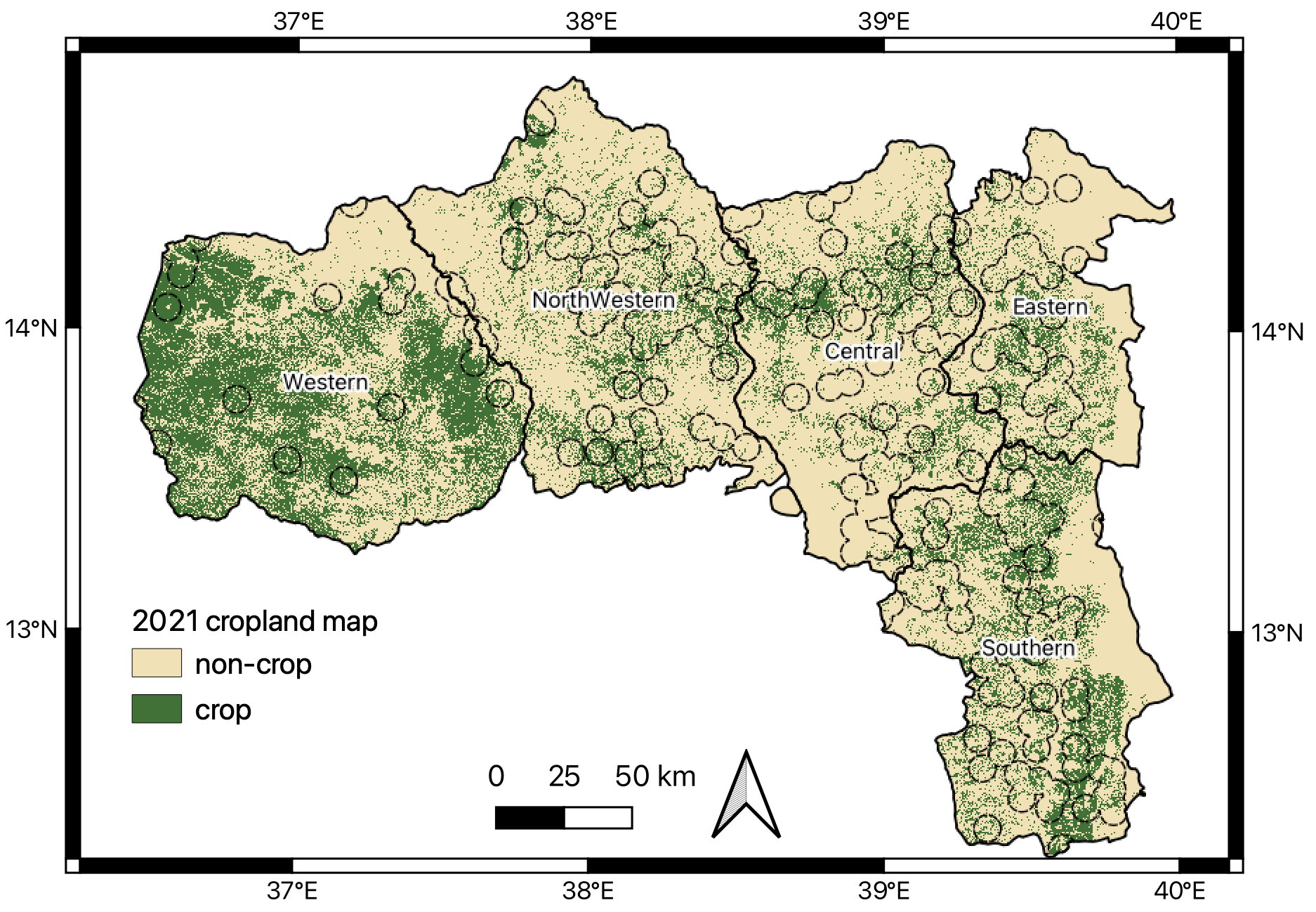}
         \caption{2021 cropland map}
         \label{fig:map2021}
     \end{subfigure}
     \begin{subfigure}[b]{0.5\textwidth}
         \centering
         \includegraphics[width=\textwidth]{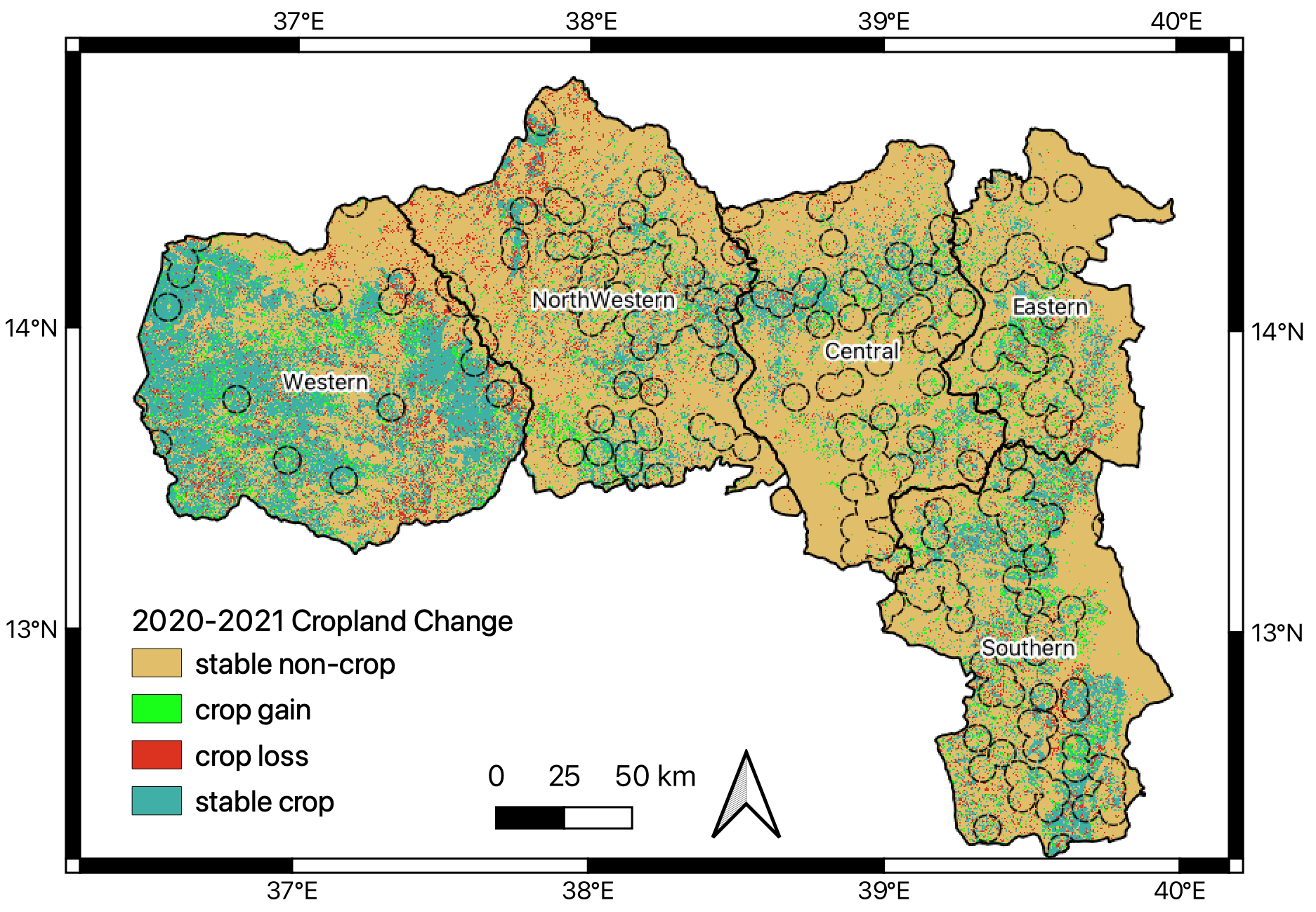}
         \caption{2020-2021 change map}
         \label{fig:changemap}
     \end{subfigure}
        \caption{Annual cropland and change maps overlaid with  zone boundaries (solid lines) and 5 km buffer around conflict events (dashed lines).}
        \label{fig:maps}
\end{figure}

Figure \ref{fig:maps} shows the annual cropland classification maps for 2020 and 2021 as well as the four-class change map of transitions between the 2020 and 2021 maps. Full resolution GeoTIFFs of maps are available at \url{https://data.harvestportal.org/dataset/annual-and-change-maps-for-tigray-2020-2021}.

Table \ref{tab:2020-accuracy} and Table \ref{tab:2021-accuracy} report the overall accuracy, precision (user's accuracy), recall (producer's accuracy), and F1 score for the cropland class in the 2020 and 2021 maps, respectively. We computed these metrics using the independent test sets labeled for each year described in Section \ref{subsec:model-datasets}. 
We found F1 scores of $0.78 \pm 0.11$ for the 2021 map and $0.69 \pm 0.11$ for 2020. Overall accuracy was similar between the two years: $0.83 \pm 0.02$ in 2020 and $0.84 \pm 0.02$ in 2021. There was a moderate difference in precision (user's accuracy) between 2020 ($0.75\pm 0.04$) and 2021 ($0.79\pm0.04$) and larger differences in recall (producer's accuracy) between 2020 ($0.65 \pm 0.03$) and 2021 ($0.76 \pm 0.03$). This is likely due to differences in climatic factors between the two years that result in different types of errors made by the classifier. For example, differences in rainfall between the two years could affect vegetation patterns that make cultivated cropland easier or harder to distinguish from fallow land or other vegetation, or differences in cloud cover could affect the quality of the input time series in some areas. Unbalanced precision and recall scores between the two years may result in higher error rates when the maps are combined to form a change map, compared to maps with balanced scores. 

\begin{table}[]
\footnotesize
    \centering
    \resizebox{\textwidth}{!}{%
    \begin{tabular}{|c|c|c|c|c|c|c|c|c|c|}
    \hline
        Map & Map Year & F1 & OA & Recall (PA) & Precision (UA) & TN & FP & FN & TP \\
    \hline \hline
    WorldCover-v100 & 2020 &	$0.70 \pm 0.11$ &	$0.84 \pm 0.02$ &	$0.60 \pm 0.03$ &	$0.84 \pm 0.04$ & 283	& 15	& 51	& 76 \\ %
    Digital Earth Africa & 2019 & $0.66 \pm 0.12$ & $0.81 \pm 0.02$ & $0.61 \pm 0.03$ & $0.72 \pm 0.04$ & 268 & 30	& 49	& 78 \\
    WorldCereal-v100 & 2020 & $0.62 \pm 0.11$ & $0.82 \pm 0.02$ & $0.49 \pm 0.03$ & $0.84 \pm 0.04$ & 286 & 12 & 65 & 62 \\
    Copernicus & 2019 & $0.53 \pm 0.12$ & $0.72 \pm 0.02$ & $0.52 \pm 0.04$ & $0.54 \pm 0.05$ & 242	& 56 & 61	& 66 \\
    GLAD &	2019 & $0.51 \pm 0.11$ & $0.78 \pm 0.02$ & $0.38 \pm 0.03$ & $0.79 \pm 0.05$ & 285	& 13	& 79	& 48 \\
    Dynamic World & 2020 &	$0.50 \pm 0.11$ &	$0.78 \pm 0.02$ &	$0.37 \pm 0.03$ &	$0.76 \pm 0.05$ &	283 & 15	& 80	& 47 \\ 
    ASAP & 2017 & $0.45 \pm 0.12$ & $0.72 \pm 0.02$ &  $0.39 \pm 0.03$ & $0.54 \pm 0.05$ & 255 & 43	& 77 & 50 \\
    Esri LULC & 2020 & $0.40 \pm 0.10$ & $0.76 \pm 0.02$ & $0.27 \pm 0.02$ & $0.77 \pm 0.06$ &	288 &	10	& 93	& 34 \\
    \hline \hline
    Harvest (ours) & 2020 & $0.69 \pm 0.11$ &	$0.83 \pm 0.02$ &	$0.65 \pm 0.03$ &	$0.75 \pm 0.04$ &	270	& 28 & 45	& 82 \\
    \hline
    \end{tabular}%
    }
    \caption{Performance metrics for 2020 test set used for assessing accuracy of annual cropland map. We report metrics for other cropland and land cover maps on the same test set for comparison to our map (ordered by decreasing F1 score). Legend: OA = Overall Accuracy, PA = Producer's Accuracy, UA = User's Accuracy, TN = True Negatives, FP = False Positives, FN = False Negatives, TP = True Positives.}
    \label{tab:2020-accuracy}
\end{table}

\begin{table}[]
    \centering
    \resizebox{\textwidth}{!}{%
    \begin{tabular}{|c|c|c|c|c|c|c|c|c|c|}
    \hline
        Map & Map Year & F1 & OA & Recall (PA) & Precision (UA) & TN & FP & FN & TP \\
    \hline \hline
    WorldCover-v200	& 2021 & $0.63 \pm 0.12$ &	$0.76 \pm 0.02$ & $0.54 \pm 0.03$ &	$0.75 \pm 0.05$ &	168 &	20 &	50 &	59  \\
    Copernicus	& 2019 & $0.57 \pm 0.09$ & $0.67 \pm 0.02$ &	 $0.53 \pm 0.02$ &	 $0.61 \pm 0.03$ &	259	 & 78 & 	108	 & 123 \\
    Digital Earth Africa & 2019 & $0.56 \pm 0.13$ &	$0.70 \pm 0.03$ &	$0.51 \pm 0.04$ &	$0.61 \pm 0.05$ &	152 & 	36 & 	53 & 	56 \\
    Dynamic World & 2021 &	$0.53 \pm 0.11$ & $0.75 \pm 0.03$ & $0.39 \pm 0.03$ & $0.83 \pm 0.05$ & 179	& 9	& 66	& 43 \\
    WorldCereal-v100 & 2020 & $0.53 \pm 0.12$ & $0.74 \pm 0.03$ & $0.40 \pm 0.03$ & $0.77 \pm 0.06$ & 175 & 13 & 65 & 44 \\
    GLAD & 2019 & $0.51 \pm 0.12$	& $0.72 \pm 0.03$	& $0.40 \pm 0.03$	& $0.71 \pm 0.06$	& 170 &	18 &	65	 & 44 \\
    ASAP & 2017 &	$0.49 \pm 0.13$ & 	$0.67 \pm 0.03$ & 	$0.42 \pm 0.04$ & 	$0.58 \pm 0.06$ & 	154	& 34	& 63	& 46 \\
    Esri LULC & 2021 & $0.47 \pm 0.12$ &	$0.72 \pm 0.03$ &	$0.34 \pm 0.03$ &	$0.77 \pm 0.06$ &	177	& 11	& 72	& 37 \\
    \hline \hline
    Harvest (ours) & 2021 & $0.78 \pm 0.11$ &	$0.84 \pm 0.02$ &	$0.76 \pm 0.03$ &	$0.79 \pm 0.04$ &	166 & 	22	& 26 & 	83 \\
    \hline
    \end{tabular}%
}
    \caption{Performance metrics for 2021 test set used for assessing accuracy of annual cropland map. We report metrics for other cropland and land cover maps on the same test set for comparison to our map (ordered by decreasing F1 score). Legend: OA = Overall Accuracy, PA = Producer's Accuracy, UA = User's Accuracy, TN = True Negatives, FP = False Positives, FN = False Negatives, TP = True Positives.}
    \label{tab:2021-accuracy}
\end{table}

Of the compared maps described in Section \ref{subsec:comparison} (not including our map), we found that WorldCover had the highest performance metrics overall for both years. Compared to our map, the WorldCover map had a similar F1 score in 2020 but had significantly lower performance in 2021. %

\begin{table}[]
    \centering
    \begin{tabular}{|c|c|c|c|c|c||c|}
        \hline
        Year & Southern & Central & Western & Eastern & North Western & Tigray \\
        \hline \hline
        2020 & $227 \pm	43$ & $256 \pm 48$ & $327 \pm 71$ & $133 \pm 37$ & $282 \pm 57$ & $1,132 \pm 133$ \\
        2021 & $265 \pm 45$ & $241 \pm 46$ & $368 \pm 76$ & $136 \pm 35$ & $314 \pm 60$ & $1,217 \pm 132$ \\
        \hline
    \end{tabular}
    \caption{Annual area estimates for Tigray and each zone within Tigray (units are kha)}
    \label{tab:annual-area}
\end{table}

\subsection{Annual area estimates}
\begin{figure}
    \centering
    \includegraphics[width=0.8\textwidth]{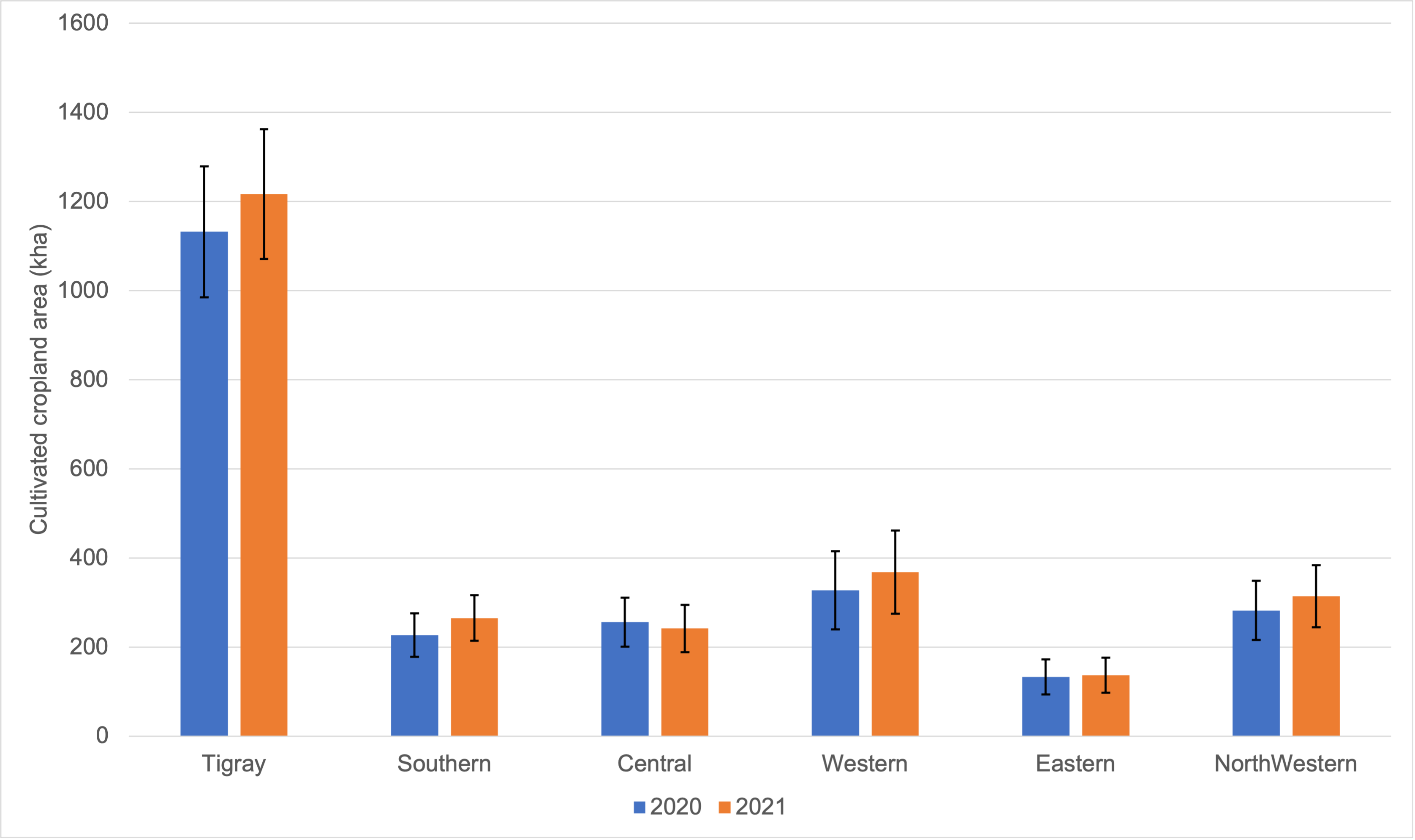}
    \caption{Annual area estimates for Tigray region and each zone within Tigray (units are kha)}
    \label{fig:annual-area}
\end{figure}

We report the annual cropland area estimates for 2020 and 2021 in Table \ref{tab:annual-area}. Figure \ref{fig:annual-area} shows these estimates as a bar plot to illustrate the relative differences between each year. 
We found similar estimates for cultivated area in Tigray in 2020 and 2021, with a possible small net increase: $1,132 \pm 133$ kha in 2020 and $1,217 \pm 132$ kha in 2021 (1 kha = 1,000 ha). We found a similar trend for each zone, with the exception of the Central zone which showed a possible small net decrease. This net decrease in Central Tigray may be due to a higher concentration of conflict events recorded in the Central zone during the growing season \citep{weldegebriel2023eyes}. Figure \ref{fig:confusion-tigray} shows the confusion matrices between the map and reference samples for Tigray. We provide the confusion matrix for each zone in Figures \ref{fig:confusion-southern}-\ref{fig:confusion-northwestern}. All confusion matrices are expressed in terms of proportion of mapped area based on good practice recommendations \citep{olofsson2014good}.

\subsection{Change area estimates}
\begin{figure}
     \centering
     \begin{subfigure}[b]{0.3\textwidth}
         \centering
         \includegraphics[width=\textwidth]{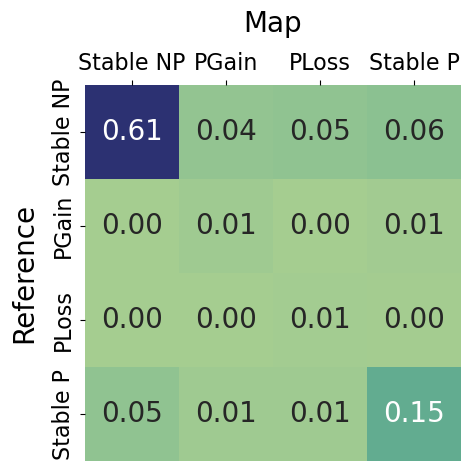}
         \caption{2020-2021 change map}
         \label{fig:cm-change-tigray}
     \end{subfigure}
     \hfill
     \begin{subfigure}[b]{0.3\textwidth}
         \centering
         \includegraphics[width=\textwidth]{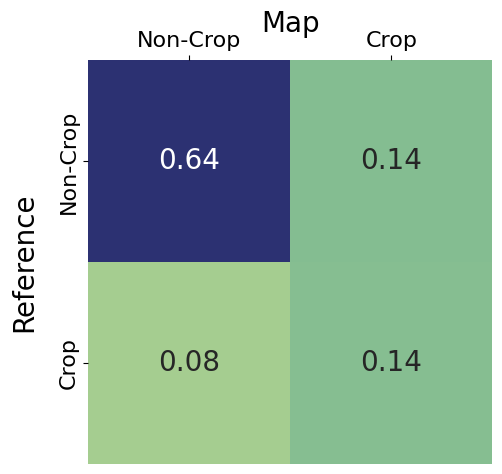}
         \caption{2020 map}
         \label{fig:cm-2020-tigray}
     \end{subfigure}
    \hfill
     \begin{subfigure}[b]{0.3\textwidth}
         \centering
         \includegraphics[width=\textwidth]{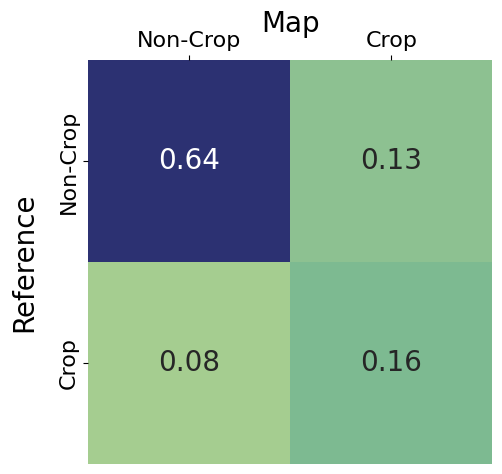}
         \caption{2021 map}
         \label{fig:cm-2021-tigray}
     \end{subfigure}
        \caption{Confusion matrices between reference samples and Tigray change and annual cropland maps, expressed in terms of proportion of area as recommended in \cite{olofsson2014good}}
        \label{fig:confusion-tigray}
\end{figure}

Table \ref{tab:change-area} summarizes the area estimates for the four transition classes. We found that cropland gain accounted for $70-160$ kha while cropland loss accounted for $19-79$ kha in Tigray. The stable transition classes accounted for the majority of the area in Tigray: $1,033-1,293$ kha of stable cropland and $3,798-4,054$ kha of stable non-cropland.
Figure \ref{fig:cm-change-tigray} shows the confusion matrix between the reference samples (stratified by the four change class strata) and the classified change map, expressed in terms of mapped area proportions. For Tigray, there was good agreement between the map and reference samples for the stable not planted class: of the area mapped as stable not planted, only a small fraction of the mapped area should have been mapped as stable planted according to the reference sample. There was also good agreement between the map and reference samples for the stable planted class, though approximately one third of the area mapped as stable planted should have been mapped as stable not planted according to the reference sample. This is likely due to the similarity in satellite time series between planted fields and other vegetated areas, which may include fallow fields in addition to non-agricultural areas. %
We used the same reference samples to compute class-wise accuracy metrics (overall accuracy, user accuracy/precision, producer accuracy/recall, true positive rate, and false positive rate) in Table \ref{tab:change-accuracy}. 

The change estimates at the zone scale follow a similar trend as the region (Tigray) scale. The majority of the area in each zone was estimated as stable non-crop followed by stable cropland. A relatively small amount of cropland gain and loss was estimated in each zone. However, the standard errors relative to the area estimates were much larger at the zone scale compared to those for Tigray region due to smaller reference sample sizes. The standard error was highest for the change classes in Western Tigray. Figures \ref{fig:cm-change-southern}-\ref{fig:cm-change-northwestern} provide the confusion matrices for each zone. Tables \ref{tab:change-accuracy-southern}-\ref{tab:change-accuracy-northwestern} report the accuracy metrics computed using the reference samples in each zone.

While we provided the full-resolution cropland and cropland change maps as part of this study for transparency and reproducibility of our area estimates, we emphasize that their intended use is for stratification in the area estimation procedure (see Figure \ref{fig:area-workflow}). The change map has limited accuracy as shown in Figures \ref{fig:confusion-southern}-\ref{fig:confusion-northwestern} and is not intended for visual interpretation of changes or other uses outside of the sample-based area estimation described in this paper. We also emphasize that inaccuracies in the map do not signify inaccuracies in the area estimates, since such errors are accounted for by the use of an unbiased estimator \cite{olofsson2020mitigating}.

\begin{table}[]
    \centering
     \resizebox{\textwidth}{!}{%
     \begin{tabular}{|c|c|c|c|c|c||c|}
        \hline
        Class & Southern & Central & Western & Eastern & North Western & Tigray \\
        \hline \hline
Stable cropland	&	$	208	\pm	43	$	&	$	226	\pm	47	$	&	$	418	\pm	80	$	&	$	108	\pm	35	$	&	$	286	\pm	58	$	&	$	1163	\pm	130	$	\\
Stable non-crop	&	$	728	\pm	47	$	&	$	775	\pm	49	$	&	$	865	\pm	78	$	&	$	473	\pm	37	$	&	$	937	\pm	59	$	&	$	3926	\pm	128	$	\\
Cropland gain	&	$	53	\pm	23	$	&	$	23	\pm	17	$	&	$	28	\pm	29	$	&	$	24	\pm	14	$	&	$	21	\pm	14	$	&	$	115	\pm	45	$	\\
Cropland loss	&	$	25	\pm	17	$	&	$	17	\pm	13	$	&	$	10	\pm	19	$	&	$	19	\pm	16	$	&	$	12	\pm	15	$	&	$	49	\pm	30	$	\\
        \hline
    \end{tabular}}
    \caption{Change area estimates for Tigray region and each zone within Tigray (units are kha)}
    \label{tab:change-area}
\end{table}

\begin{table}[]
    \centering
   \resizebox{\textwidth}{!}{%
   \begin{tabular}{|c|c|c|c|c|}
        \hline
        Metric & Precision (UA) & Recall (PA) & True Positive Rate & False Positive Rate \\
        \hline \hline
        Stable cropland & $0.69 \pm 0.08$ & $0.67 \pm 0.06$ & $0.53$ & $0.07$\\
        Stable non-crop & $0.93 \pm 0.03$ & $0.81 \pm 0.02$ & $0.65$ & $0.13$  \\
        Cropland gain & $0.19 \pm 0.07$ & $0.55 \pm 0.21$ & $0.76$ & $0.14$  \\
        Cropland loss & $0.08 \pm 0.05$ & $0.59 \pm 0.33$ & $0.79$ & $0.15$ \\
        \hline
    \end{tabular}}
    \caption{Accuracy metrics for four-class cropland change reference sample in Tigray. Overall accuracy is $0.77 \pm 0.02$. See Tables \ref{tab:change-accuracy-southern}-\ref{tab:change-accuracy-northwestern} for zone-level metrics. Legend: OA = Overall Accuracy, PA = Producer's Accuracy, UA = User's Accuracy, TN = True Negatives, FP = False Positives, FN = False Negatives, TP = True Positives.} 
    \label{tab:change-accuracy}
\end{table}

\subsection{Cropland area changes in proximity to conflict events}
Table \ref{tab:change-area-buffer} summarizes the change area estimates computed inside and outside of the 5 km conflict buffer. We reported the confidence interval in units of kha and as a percentage of the total area in each boundary to enable comparison between the two boundaries, since the total area outside the buffer is much larger than the area inside the buffer. We found that the percentage of total area estimated for each change class was similar within the confidence intervals of the estimates. However, our results show a higher upper boundary of cropland loss inside the conflict event buffer ($0-3\%$) compared to outside the buffer ($0-1\%$). This suggests that cropland loss may have been slightly more concentrated in the vicinity of conflict events compared to locations with greater distance from conflict events. Figure \ref{fig:confusion-buffer} shows the confusion matrix between the reference samples and the mapped areas inside and outside of the conflict event buffer. Tables \ref{tab:change-accuracy-buffer5km} and \ref{tab:change-accuracy-outsidebuffer} report the accuracy metrics.

Since the reference sample size for the area inside the conflict buffer ($n=219$) was smaller than the sample size for the area outside the conflict buffer ($n=582$), the higher upper bound of loss in the conflict buffer could be due to a relatively smaller sample size. To test this, we performed an experiment to estimate the cropland loss area outside the conflict buffer with a smaller sample size equal to that inside the conflict buffer. We randomly sub-sampled 219 points from the non-conflict 582 reference samples, then calculated the cropland loss area and confidence interval (CI) using this sample. We repeated this 10 times using different random seeds, since the result may vary depending on the samples drawn. We calculated the median and mean cropland loss area and CI over all random seeds. Table \ref{tab:conflict-buffer-seeds} reports the results from this experiment and Table \ref{tab:conflict-buffer-sample-test} summarizes the results compared to those reported in Table \ref{tab:change-area-buffer} using the full reference sample. For both the mean and median estimates with $n=219$, we found a lower upper bound CI as a percentage of total area ($0-1\%$ for the median, $0-2\%$ for the mean), consistent with our result using the larger sample size ($0-1\%$). We note that the mean estimate is sensitive to outliers thus we feel the median is a more representative estimate. These results confirm that the higher upper bound of cropland loss inside the 5 km conflict buffer is most likely due to a higher likelihood of loss near conflict events, rather than a smaller sample size.

\begin{table}[]
    \centering
    \begin{tabular}{|c|c|c|}
        \hline
        Class & Inside Buffer & Outside Buffer \\
        \hline \hline
Stable cropland	& $318	\pm	63	$ ($	19	-	29 \%	$)	&	$	838	\pm	115	$ ($	18	-	24 \%	$)	\\
Stable non-crop	& $	943	\pm	62	$ ($	67	-	76 \%	$)	&	$	2999	\pm	113	$ ($	73	-	79 \%	$)	\\
Cropland gain	& $	37	\pm	19	$ ($	1	-	4 \%	$)	&	$	77	\pm	42	$ ($	1	-	3 \%	$)	\\
Cropland loss	&	$	19	\pm	17	$ ($	0	-	3 \%	$)	&	$	28	\pm	23	$ ($	0	-	1 \%	$)	\\
        \hline
    \end{tabular}
    \caption{Change area estimates (kha) inside and outside of a 5 km buffer around conflict events.}
    \label{tab:change-area-buffer}
\end{table}

\section{Discussion}
\label{sec:discussion}

\subsection{Key findings and agreement with other studies}
In reports published in the first half of 2021 prior to the planting and growing season, there were grave concerns and uncertainty about whether Tigrayan farmers would be able to cultivate their crops in the 2021 meher season \citep{un-report,nyssen2023crop}.
Satellite remote sensing data provides real-time and historical observations that can be used for qualitative and quantitative assessments of how conflicts affect agricultural land use and production. This source of data is particularly valuable for evaluating impacts during armed conflict, since traditional ground-based surveys may be impractical due to safety concerns.
Using remote sensing observations of Tigray during and prior to the war, studies such as ours, Nyssen et al.~\citep{nyssen2023crop}, and Weldegebriel et al.~\citep{weldegebriel2023eyes} could ascertain information to help clarify uncertainty about the impact of the conflict on cultivated area.

Our finding that crop cultivation area in Tigray in wartime 2021 was comparable to cultivated area in pre-war 2020 agrees with previous studies of the impact of the war on cultivated area \citep{nyssen2023crop,weldegebriel2023eyes}. Through ground interviews and satellite image analysis, these studies concluded that despite many reports of disruption to farming activities, most Tigrayan farmers were ultimately able to plant and harvest crops by employing wartime coping strategies such as shifting the timing and location of farming activities, changing crop types, and leveraging communal aid \citep{nyssen2023crop}. While we did not find a net loss of cropland as in Weldegebriel et al.~\citep{weldegebriel2023eyes}, it is difficult to directly compare the 714 km$^2$ net loss found by their study to our study since their study was restricted to highland croplands in Tigray,  employed a specific definition of ``well-cultivated cropland,'' estimated area using pixel counting instead of a reference sample, and did not provide confidence intervals. Nevertheless, our estimate of area attributed to cropland loss in Tigray was between 19-79 kha which is comparable to the net loss of 714 km$^2$ (or 71.4 kha) found by Weldegebriel et al.~\citep{weldegebriel2023eyes}. 

Our estimate of cropland loss in Tigray is much lower than cropland losses found in studies of some other conflict regions globally, which have been found to range from 15-30\% \citep{skakun2019satellite,witmer2008detecting,boudinaud-mali}. Our findings further support the conclusion by Nyssen et al.~\citep{nyssen2023crop} that Tigrayan smallholder farmers were largely able to sustain cultivation amidst violence. However, this may to some extent also reflect the difficulty of quantifying changes in small, heterogeneous farm landscapes like Tigray and confusion between planted and fallow fields, as discussed in Section \ref{sec:limitations} and Yin et al.~\citep{yin2020monitoring}.

\subsection{Limitations}
\label{sec:limitations}
We followed best practices established by the scientific literature whenever possible, but there are some limitations inherent in our approach.
One possible source of error in our estimates is in the reference labels used to estimate area. We tried to maximize the correctness and confidence of these reference labels as much as possible through our response design. We used high-resolution images from PlanetScope (3 m/pixel) and Google Earth ($<$1 m/pixel) when available. Each sample was analyzed and labeled independently by at least two different annotators. If there was any disagreement between annotators, we convened expert annotators to review and discuss all of the available evidence and come to a consensus label for the sample. Even with these precautions, it is possible that some samples were interpreted incorrectly due to the difficulty of differentiating between crop fields that are fallow and fields that are planted from satellite images, even in high-resolution (3 m/pixel) images (see Figure \ref{fig:ghent-data}). \ref{sec:reference-examples} shows some examples of reference sample labels and corresponding high-resolution satellite images from the reference dataset. 

Another limitation of this study is that we used only two years for our change analysis: 2021 (wartime) and 2020 (pre-war). Including additional pre-war years in the analysis would help establish a baseline of ``nominal'' area estimates to compare with wartime 2021. It is also possible that in addition to the outbreak of war during the end of the harvest period in 2020, the COVID-19 pandemic may have had some impact on the 2020 growing season \citep{okolie2022effect}. Nevertheless, we included only 2020 and 2021 in our study due to the significant resources required for analyzing two years and the precedence set by prior studies using similar statistical techniques that compared two years for change assessments (e.g., \citep{olofsson2014good,skakun2019satellite,olsen2021impact}).

Finally, the impacts of the Tigray war on crop cultivation in our study were limited to those that resulted in a binary change in cultivation status: either cultivated or not cultivated. We designed our study this way because our primary goal was to provide early warning insights to the USAID Famine Early Warning Systems Network's Early Warning Team about the potential for shortfalls in crop production in Tigray during the war. Our goal was to broadly assess whether the total cropland area under cultivation in Tigray was lower during the war (2021) compared to prior years (2020). Thus, we followed a similar approach to previous work that estimated the change in cropland area between a pre-war and wartime period (e.g., \cite{skakun2019satellite,boudinaud-mali,ma2022spatiotemporal,eklund2017conflict,witmer2008detecting}).

Our study did not account for other impacts such as population displacement, changes in crop types or cultivation practices, movement and access limitations, market dysfunction and loss of income, or destruction of food stocks and farming equipment that were reported in farmer surveys \citep{un-report,nyssen2023crop}. While some studies analyzed plowing status and harvest status from satellite observations \citep{nyssen2023crop,weldegebriel2023eyes}, most of the other factors listed are difficult or impossible to determine from satellite observations, especially in the absence of extensive ground-truth data for calibration. Other studies used quasi-experimental methods such as propensity score matching to account for additional covariates such as population density, distance to roads, or distance to conflict events \citep{olsen2021impact}. We did not employ such techniques since the objective of our study was a broad assessment of cropland area change occurring between the pre-war (2020) and wartime (2021) period. In addition, as shown in Figure \ref{fig:acled-map}, conflict events during the study period were widely distributed and not aggregated along a front or in clusters. Using the distance to conflict or the density of conflicts as a predictor variable may not adequately capture the nuanced impacts that individual conflict events can have on cropland loss due to the heterogeneous nature of warfare tactics, varying levels of intensity, and the different scales at which these events occur.  

\section{Conclusions}
\label{sec:conclusion}
Despite widespread disruptions to agriculture, our remote sensing-based crop maps and area estimates suggest Tigrayan smallholder farmers were largely able to carry out crop cultivation activities in 2021 amidst the Tigray war. Our approach using satellite remote sensing, machine learning, and statistical techniques enables timely, accurate insights about agriculture in conflict regions where ground access is not possible using satellite data and statistical methods. Further studies could build on our methodology to incorporate additional years before and during the Tigray war to continue monitoring Tigray's agriculture and food security situation. We made all code and data used in this study publicly accessible to enable transparency and reproducibility and to facilitate future work.

\section*{Acknowledgments}
This work was supported by NASA under Award \#80NSSC21K1511 and the NASA Harvest Consortium on Food Security and Agriculture (Award \#80NSSC18M0039). We thank the Famine Early Warning Systems Network (FEWS NET) Science and Early Warning Teams for assistance with point labeling and discussions on early versions of results. We thank Liya Weldegebriel, Emnet Negash, and Jan Nyssen for sharing their concurrent findings, local expertise, and ground-truth data which helped strengthen this study. Finally, we thank the anonymous journal reviewers for their constructive feedback that greatly improved this study.

\bibliographystyle{elsarticle-num}
\bibliography{references.bib}

\begin{thebibliography}{10}
\expandafter\ifx\csname url\endcsname\relax
  \def\url#1{\texttt{#1}}\fi
\expandafter\ifx\csname urlprefix\endcsname\relax\def\urlprefix{URL }\fi
\expandafter\ifx\csname href\endcsname\relax
  \def\href#1#2{#2} \def\path#1{#1}\fi

\bibitem{morton2007impact}
J.~F. Morton, The impact of climate change on smallholder and subsistence
  agriculture, Proceedings of the national academy of sciences 104~(50) (2007)
  19680--19685.

\bibitem{flores2005food}
M.~Flores, Y.~Khwaja, P.~White, Food security in protracted crises: building
  more effective policy frameworks, Disasters 29 (2005) S25--S51.

\bibitem{nyssen2023crop}
J.~Nyssen, E.~Negash, B.~Van~Schaeybroeck, K.~Haegeman, S.~Annys, Crop
  cultivation at wartime--plight and resilience of {T}igray’s agrarian
  society ({N}orth {E}thiopia), Defence and Peace Economics 34~(5) (2023)
  618--645.

\bibitem{suhrke1993pressure}
A.~Suhrke, S.~Hazarika, Pressure points: environmental degradation, migration
  and conflict (1993).

\bibitem{morales2022co}
H.~Morales-Mu{\~n}oz, A.~Bailey, K.~L{\"o}hr, G.~Caroli, M.~E.~J. Villarino,
  A.~M. LoboGuerrero, M.~Bonatti, S.~Siebert, A.~Castro-Nu{\~n}ez, Co-benefits
  through coordination of climate action and peacebuilding: a system dynamics
  model, Journal of Peacebuilding \& Development 17~(3) (2022) 304--323.

\bibitem{skakun2019satellite}
S.~Skakun, C.~O. Justice, N.~Kussul, A.~Shelestov, M.~Lavreniuk, Satellite data
  reveal cropland losses in south-eastern {U}kraine under military conflict,
  Frontiers in Earth Science (2019) 305.

\bibitem{ma2022spatiotemporal}
Y.~Ma, D.~Lyu, K.~Sun, S.~Li, B.~Zhu, R.~Zhao, M.~Zheng, K.~Song,
  Spatiotemporal analysis and war impact assessment of agricultural land in
  {U}kraine using {RS} and {GIS} technology, Land 11~(10) (2022) 1810.

\bibitem{witmer2008detecting}
F.~D. Witmer, Detecting war-induced abandoned agricultural land in northeast
  {B}osnia using multispectral, multitemporal {Landsat TM} imagery,
  International Journal of Remote Sensing 29~(13) (2008) 3805--3831.

\bibitem{boudinaud-mali}
L.~Boudinaud, S.~A. Orenstein,
  \href{https://isprs-archives.copernicus.org/articles/XLVI-4-W2-2021/9/2021/}{Assessing
  cropland abandonment from violent conflict in central mali with sentinel-2
  and google earth engine}, The International Archives of the Photogrammetry,
  Remote Sensing and Spatial Information Sciences XLVI-4/W2-2021 (2021) 9--15.
\newblock \href {https://doi.org/10.5194/isprs-archives-XLVI-4-W2-2021-9-2021}
  {\path{doi:10.5194/isprs-archives-XLVI-4-W2-2021-9-2021}}.
\newline\urlprefix\url{https://isprs-archives.copernicus.org/articles/XLVI-4-W2-2021/9/2021/}

\bibitem{olsen2021impact}
V.~M. Olsen, R.~Fensholt, P.~Olofsson, R.~Bonifacio, V.~Butsic, D.~Druce,
  D.~Ray, A.~V. Prishchepov, The impact of conflict-driven cropland abandonment
  on food insecurity in {South Sudan} revealed using satellite remote sensing,
  Nature Food 2~(12) (2021) 990--996.

\bibitem{yin2019agricultural}
H.~Yin, V.~Butsic, J.~Buchner, T.~Kuemmerle, A.~V. Prishchepov, M.~Baumann,
  E.~V. Bragina, H.~Sayadyan, V.~C. Radeloff, Agricultural abandonment and
  re-cultivation during and after the {Chechen Wars} in the northern caucasus,
  Global Environmental Change 55 (2019) 149--159.

\bibitem{muller2009causes}
D.~M{ü}ller, T.~Kuemmerle, Causes of cropland abandonment during the
  post-socialist transition in southern romania, in: Regional aspects of
  climate-terrestrial-hydrologic interactions in non-boreal Eastern Europe,
  Springer, 2009, pp. 221--230.

\bibitem{peterson1998changes}
U.~Peterson, R.~Aunap, Changes in agricultural land use in {E}stonia in the
  1990s detected with multitemporal {Landsat MSS} imagery, Landscape and urban
  planning 41~(3-4) (1998) 193--201.

\bibitem{li2022civil}
X.-Y. Li, X.~Li, Z.~Fan, L.~Mi, T.~Kandakji, Z.~Song, D.~Li, X.-P. Song, Civil
  war hinders crop production and threatens food security in {S}yria, Nature
  Food 3~(1) (2022) 38--46.

\bibitem{eklund2017conflict}
L.~Eklund, M.~Degerald, M.~Brandt, A.~V. Prishchepov, P.~Pilesj{\"o}, How
  conflict affects land use: agricultural activity in areas seized by the
  islamic state, Environmental Research Letters 12~(5) (2017) 054004.

\bibitem{harvest-ukraine}
{NASA Harvest},
  \href{https://nasaharvest.org/news/farming-warzone-nasa-harvest-releases-satellite-based-ukraine-wheat-production-estimates}{{Farming
  A Warzone: NASA Harvest Releases Satellite-Based Ukraine Wheat Production
  Estimates}} (June 2023).
\newline\urlprefix\url{https://nasaharvest.org/news/farming-warzone-nasa-harvest-releases-satellite-based-ukraine-wheat-production-estimates}

\bibitem{yin2020monitoring}
H.~Yin, A.~Brand{\~a}o~Jr, J.~Buchner, D.~Helmers, B.~G. Iuliano, N.~E.
  Kimambo, K.~E. Lewi{\'n}ska, E.~Razenkova, A.~Rizayeva, N.~Rogova, et~al.,
  Monitoring cropland abandonment with {L}andsat time series, Remote Sensing of
  Environment 246 (2020) 111873.

\bibitem{un-report}
{Integrated Food Security Phase Classification},
  \href{https://www.ipcinfo.org/fileadmin/user_upload/ipcinfo/docs/IPC_Ethiopia_Acute_Food_Insecurity_2021MaySept_national.pdf}{{IPC
  Acute Food Insecurity Analysis}} (June 2021).
\newline\urlprefix\url{https://www.ipcinfo.org/fileadmin/user_upload/ipcinfo/docs/IPC_Ethiopia_Acute_Food_Insecurity_2021MaySept_national.pdf}

\bibitem{weldegebriel2023eyes}
L.~Weldegebriel, E.~Negash, J.~Nyssen, D.~Lobell,
  \href{https://doi.org/10.21203/rs.3.rs-2649237/v1}{Eyes in the sky on
  {T}igray - monitoring the impact of armed conflict on cultivated land using
  satellite imagery in {E}thiopia} (2023).
\newline\urlprefix\url{https://doi.org/10.21203/rs.3.rs-2649237/v1}

\bibitem{olofsson2014good}
P.~Olofsson, G.~M. Foody, M.~Herold, S.~V. Stehman, C.~E. Woodcock, M.~A.
  Wulder, Good practices for estimating area and assessing accuracy of land
  change, Remote sensing of Environment 148 (2014) 42--57.

\bibitem{annys2021tigray}
S.~Annys, T.~Vandenbrempt, E.~Negash, L.~De~Sloover, J.~Nyssen, Tigray: atlas
  of the humanitarian situation (2021).

\bibitem{CSAEthiopia2017}
{Central Statistical Agency Ethiopia}, {ICF}, {Demographic and Health Survey
  2016--ETHIOPIA: Demographic and Health Survey 2016} (July 2017).

\bibitem{Ethiopia2015AgriculturalSurvey}
{Central Statistical Agency}, The federal democratic republic of ethiopia
  central statistical agency agricultural sample survey 2014/2015 (2007 e.c.),
  Agricultural sample survey, The Federal Democratic Republic of Ethiopia
  Central Statistical Agency (2015).

\bibitem{FEWSNET2021EthiopiaAlert}
{Famine Early Warning Systems Network (FEWS NET)},
  \href{https://fews.net/sites/default/files/documents/reports/Ethiopia-alert-20210517.pdf}{Ethiopia
  food security alert}, Alert report, Famine Early Warning Systems Network
  (FEWS NET), accessed: 2024-04-06 (5 2021).
\newline\urlprefix\url{https://fews.net/sites/default/files/documents/reports/Ethiopia-alert-20210517.pdf}

\bibitem{rapidresponse}
H.~Kerner, G.~Tseng, I.~Becker-Reshef, C.~Nakalembe, B.~Barker, B.~Munshell,
  M.~Paliyam, M.~Hosseini,
  \href{https://doi.org/10.48550/arXiv.2006.16866}{Rapid response crop maps in
  data sparse regions}, in: ACM SIGKDD Conference on Data Mining and Knowledge
  Discovery Workshops, 2020, pp. 1--7.
\newline\urlprefix\url{https://doi.org/10.48550/arXiv.2006.16866}

\bibitem{openmapflow}
I.~Zvonkov, G.~Tseng, C.~Nakalembe, H.~Kerner,
  \href{https://ojs.aaai.org/index.php/AAAI/article/view/26713}{{OpenMapFlow}:
  A library for rapid map creation with machine learning and remote sensing
  data}, Proceedings of the AAAI Conference on Artificial Intelligence 37~(12)
  (2023) 14655--14663.
\newblock \href {https://doi.org/10.1609/aaai.v37i12.26713}
  {\path{doi:10.1609/aaai.v37i12.26713}}.
\newline\urlprefix\url{https://ojs.aaai.org/index.php/AAAI/article/view/26713}

\bibitem{cropharvest}
G.~Tseng, I.~Zvonkov, C.~Nakalembe, H.~Kerner,
  \href{https://datasets-benchmarks-proceedings.neurips.cc/paper_files/paper/2021/file/54229abfcfa5649e7003b83dd4755294-Paper-round2.pdf}{Crop{H}arvest:
  a global satellite dataset for crop type classification}, in: Neural
  Information Processing Systems (NeurIPS) Datasets and Benchmarks Track, 2021,
  pp. 1--14.
\newline\urlprefix\url{https://datasets-benchmarks-proceedings.neurips.cc/paper_files/paper/2021/file/54229abfcfa5649e7003b83dd4755294-Paper-round2.pdf}

\bibitem{cloudfree}
M.~Schmitt, L.~Hughes, C.~Qiu, X.~Zhu, Aggregating cloud-free {S}entinel-2
  images with {Google Earth En}gine, ISPRS Annals of Photogrammetry, Remote
  Sensing and Spatial Information Sciences IV-2/W7 (2019) 145--152.
\newblock \href {https://doi.org/10.5194/isprs-annals-IV-2-W7-145-2019}
  {\path{doi:10.5194/isprs-annals-IV-2-W7-145-2019}}.

\bibitem{ghentdata2021}
T.~Asfaha, J.~Nyssen, E.~Negash, H.~Meaza, Z.~Tesfamariam, Spatially explicit
  dataset on crop status of 161 farm plots in tigray (20-30 august 2021) (04
  2022).
\newblock \href {https://doi.org/10.1594/PANGAEA.943374}
  {\path{doi:10.1594/PANGAEA.943374}}.

\bibitem{zanaga2021}
D.~Zanaga, R.~Van De~Kerchove, W.~De~Keersmaecker, N.~Souverijns, C.~Brockmann,
  R.~Quast, J.~Wevers, A.~Grosu, A.~Paccini, S.~Vergnaud, O.~Cartus,
  M.~Santoro, S.~Fritz, I.~Georgieva, M.~Lesiv, S.~Carter, M.~Herold, L.~Li,
  N.-E. Tsendbazar, F.~Ramoino, O.~Arino,
  \href{https://doi.org/10.5281/zenodo.5571936}{{ESA WorldCover} 10 m 2020
  v100} (Oct. 2021).
\newblock \href {https://doi.org/10.5281/zenodo.5571936}
  {\path{doi:10.5281/zenodo.5571936}}.
\newline\urlprefix\url{https://doi.org/10.5281/zenodo.5571936}

\bibitem{kerner2023accurate}
H.~Kerner, C.~Nakalembe, A.~Yang, I.~Zvonkov, R.~McWeeny, G.~Tseng,
  I.~Becker-Reshef, \href{https://arxiv.org/abs/2307.02575}{How accurate are
  existing land cover maps for agriculture in sub-saharan africa?}, arXiv
  preprint arXiv:2307.02575 (2023).
\newline\urlprefix\url{https://arxiv.org/abs/2307.02575}

\bibitem{zanaga2022esa}
D.~Zanaga, R.~Van De~Kerchove, D.~Daems, W.~De~Keersmaecker, C.~Brockmann,
  G.~Kirches, J.~Wevers, O.~Cartus, M.~Santoro, S.~Fritz, et~al., {ESA
  WorldCover} 10 m 2021 v200 (2022).

\bibitem{worldcover-data}
{ESA WorldCover consortium}, {ESA WorldCover project} (2020).
\newblock \href {https://doi.org/https://esa-worldcover.org/en/data-access}
  {\path{doi:https://esa-worldcover.org/en/data-access}}.

\bibitem{burton2022co}
C.~Burton, F.~Yuan, C.~Ee-Faye, M.~Halabisky, D.~Ongo, F.~Mar, V.~Addabor,
  B.~Mamane, S.~Adimou,
  \href{https://www.researchgate.net/publication/357753449_Co-Production_of_a_10m_Cropland_Extent_Map_for_Continental_Africa_using_Sentinel-2_Cloud_Computing_and_the_Open_Data_Cube}{Co-production
  of a 10-m cropland extent map for continental {A}frica using {S}entinel-2,
  cloud computing, and the open-data-cube}, Authorea Preprints (2022).
\newline\urlprefix\url{https://www.researchgate.net/publication/357753449_Co-Production_of_a_10m_Cropland_Extent_Map_for_Continental_Africa_using_Sentinel-2_Cloud_Computing_and_the_Open_Data_Cube}

\bibitem{dea-data}
{Digital Earth Africa}, {Digital Earth Africa Cropland Extent Map (2019)}
  (2019).
\newblock \href
  {https://doi.org/https://registry.opendata.aws/deafrica-crop-extent/}
  {\path{doi:https://registry.opendata.aws/deafrica-crop-extent/}}.

\bibitem{buchhorn2020copernicus}
M.~Buchhorn, M.~Lesiv, N.-E. Tsendbazar, M.~Herold, L.~Bertels, B.~Smets,
  Copernicus global land cover layers—collection 2, Remote Sensing 12~(6)
  (2020) 1044.

\bibitem{cgls-data}
M.~Buchhorn, B.~Smets, L.~Bertels, B.~D. Roo, M.~Lesiv, N.-E. Tsendbazar,
  M.~Herold, S.~Fritz,
  \href{https://doi.org/10.5281/zenodo.3939050}{{Copernicus Global Land
  Service: Land Cover 100m: collection 3: epoch 2019: Globe}} (Sep. 2020).
\newblock \href {https://doi.org/10.5281/zenodo.3939050}
  {\path{doi:10.5281/zenodo.3939050}}.
\newline\urlprefix\url{https://doi.org/10.5281/zenodo.3939050}

\bibitem{brown2022dynamic}
C.~F. Brown, S.~P. Brumby, B.~Guzder-Williams, T.~Birch, S.~B. Hyde,
  J.~Mazzariello, W.~Czerwinski, V.~J. Pasquarella, R.~Haertel,
  S.~Ilyushchenko, et~al., Dynamic world, near real-time global 10 m land use
  land cover mapping, Scientific Data 9~(1) (2022) 251.

\bibitem{rembold2019asap}
F.~Rembold, M.~Meroni, F.~Urbano, G.~Csak, H.~Kerdiles, A.~Perez-Hoyos,
  G.~Lemoine, O.~Leo, T.~Negre, {ASAP}: A new global early warning system to
  detect anomaly hot spots of agricultural production for food security
  analysis, Agricultural systems 168 (2019) 247--257.

\bibitem{asap-data}
A.~Perez-Hoyos, Global crop and rangeland masks [data set] (2010).
\newblock \href {https://doi.org/http://data.europa.eu/89h/jrc-10112-10005}
  {\path{doi:http://data.europa.eu/89h/jrc-10112-10005}}.

\bibitem{potapov2022global}
P.~Potapov, S.~Turubanova, M.~C. Hansen, A.~Tyukavina, V.~Zalles, A.~Khan,
  X.-P. Song, A.~Pickens, Q.~Shen, J.~Cortez, Global maps of cropland extent
  and change show accelerated cropland expansion in the twenty-first century,
  Nature Food 3~(1) (2022) 19--28.

\bibitem{glad-data}
P.~Potapov, S.~Turubanova, M.~Hansen, A.~Tyukavina, V.~Zalles, A.~Khan, X.-P.
  Song, A.~Pickens, Q.~Shen, J.~Cortez., Global cropland expansion in the 21st
  century (2021).
\newblock \href {https://doi.org/https://glad.umd.edu/dataset/croplands}
  {\path{doi:https://glad.umd.edu/dataset/croplands}}.

\bibitem{karra2021global}
K.~Karra, C.~Kontgis, Z.~Statman-Weil, J.~C. Mazzariello, M.~Mathis, S.~P.
  Brumby, Global land use/land cover with {S}entinel 2 and deep learning, in:
  2021 IEEE international geoscience and remote sensing symposium IGARSS, IEEE,
  2021, pp. 4704--4707.

\bibitem{esri-data}
{Esri}, {Esri Sentinel-2 10m Land Use/Land Cover} (2022).
\newblock \href
  {https://doi.org/https://livingatlas.arcgis.com/landcoverexplorer/}
  {\path{doi:https://livingatlas.arcgis.com/landcoverexplorer/}}.

\bibitem{olofsson2020mitigating}
P.~Olofsson, P.~Ar{\'e}valo, A.~B. Espejo, C.~Green, E.~Lindquist, R.~E.
  McRoberts, M.~J. Sanz, Mitigating the effects of omission errors on area and
  area change estimates, Remote Sensing of Environment 236 (2020) 111492.

\bibitem{xu2024harvestnet}
J.~Xu, A.~Elmustafa, L.~Weldegebriel, E.~Negash, R.~Lee, C.~Meng, S.~Ermon,
  D.~Lobell, Harvestnet: A dataset for detecting smallholder farming activity
  using harvest piles and remote sensing, in: Proceedings of the AAAI
  Conference on Artificial Intelligence, Vol.~38, 2024, pp. 22438--22446.

\bibitem{okolie2022effect}
C.~C. Okolie, A.~A. Ogundeji, Effect of {COVID}-19 on agricultural production
  and food security: A scientometric analysis, Humanities and Social Sciences
  Communications 9~(1) (2022).

\end{thebibliography}

\newpage

\appendix

\section{Examples of reference sample interpretations from satellite images}
\label{sec:reference-examples}
In this section, we provide several examples of reference data samples to illustrate how we interpreted change class labels for each point based on high-resolution satellite images from Planet Monthly Mosaics and Google Earth points. In the caption of each figure, we provide the ID of the point in \texttt{change\_2020-2021\_strat\_ref\_samples\_labeled.shp} file provided at \url{https://data.harvestportal.org/dataset/annual-and-change-maps-for-tigray-2020-2021}. We indicate the approximate location of each point within Tigray by the gold diamond on each figure. Red dots indicate the point was labeled as non-crop/not cultivated while green dots indicate the point was labeled as crop/cultivated.

\begin{figure}[ht]
    \centering
    \includegraphics[width=\textwidth]{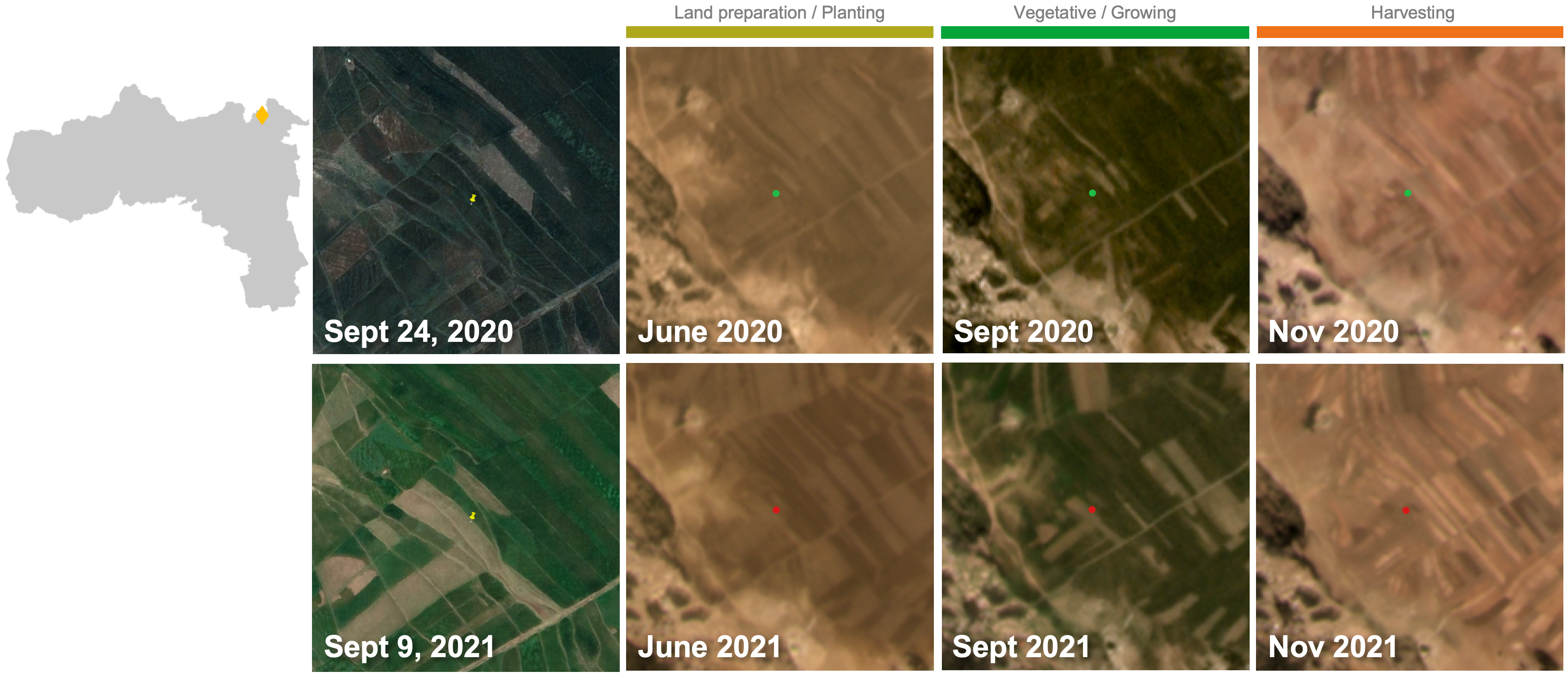}
    \caption{Example point labeled as planted cropland loss (id \#238). The reference point is in a field that appears cultivated in 2020 but fallow in 2021, although there are adjacent fields that are cultivated in 2021. The first column shows images from Airbus/CNES from Google Earth while other columns show Planet Monthly mosaics from Collect Earth Online.}
    \label{fig:p238-change-loss}
\end{figure}

\begin{figure}[ht]
    \centering
    \includegraphics[width=\textwidth]{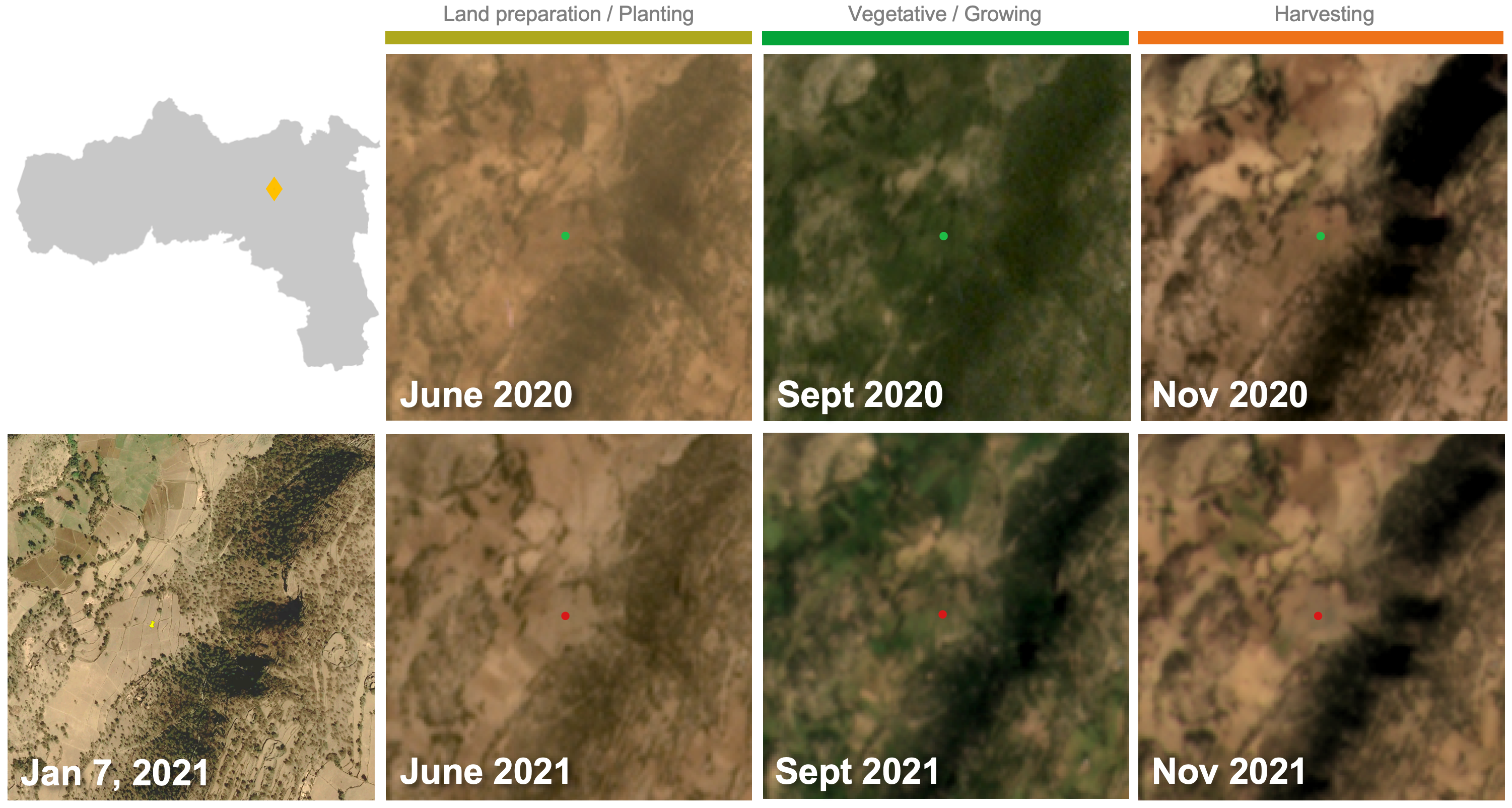}
    \caption{Example point labeled as planted cropland loss (id \#485). The reference point is in a field that appears cultivated in 2020 but fallow in 2021. In June 2020, the field containing the point appears dark brown suggesting land preparation; in September, it appears dark green and uniformly vegetated; in November, it appears to have been cleared/harvested. However, in 2021, nearby fields show signs of preparation with dark brown soil in June but not the reference field; in September, the field has some faint greenness but is sparse and patchy; in November, the field appears to have sparse and lingering vegetation indicative of a fallow field. The first column shows an image from Airbus/CNES from Google Earth (no high-resolution image was available for 2020) while other columns show Planet Monthly mosaics from Collect Earth Online. The Google Earth image was taken outside of the growing season, but still provides helpful context showing the area in higher resolution than can be seen in the Planet mosaics.}
    \label{fig:p485-change-loss}
\end{figure}

\begin{figure}[ht]
    \centering
    \includegraphics[width=\textwidth]{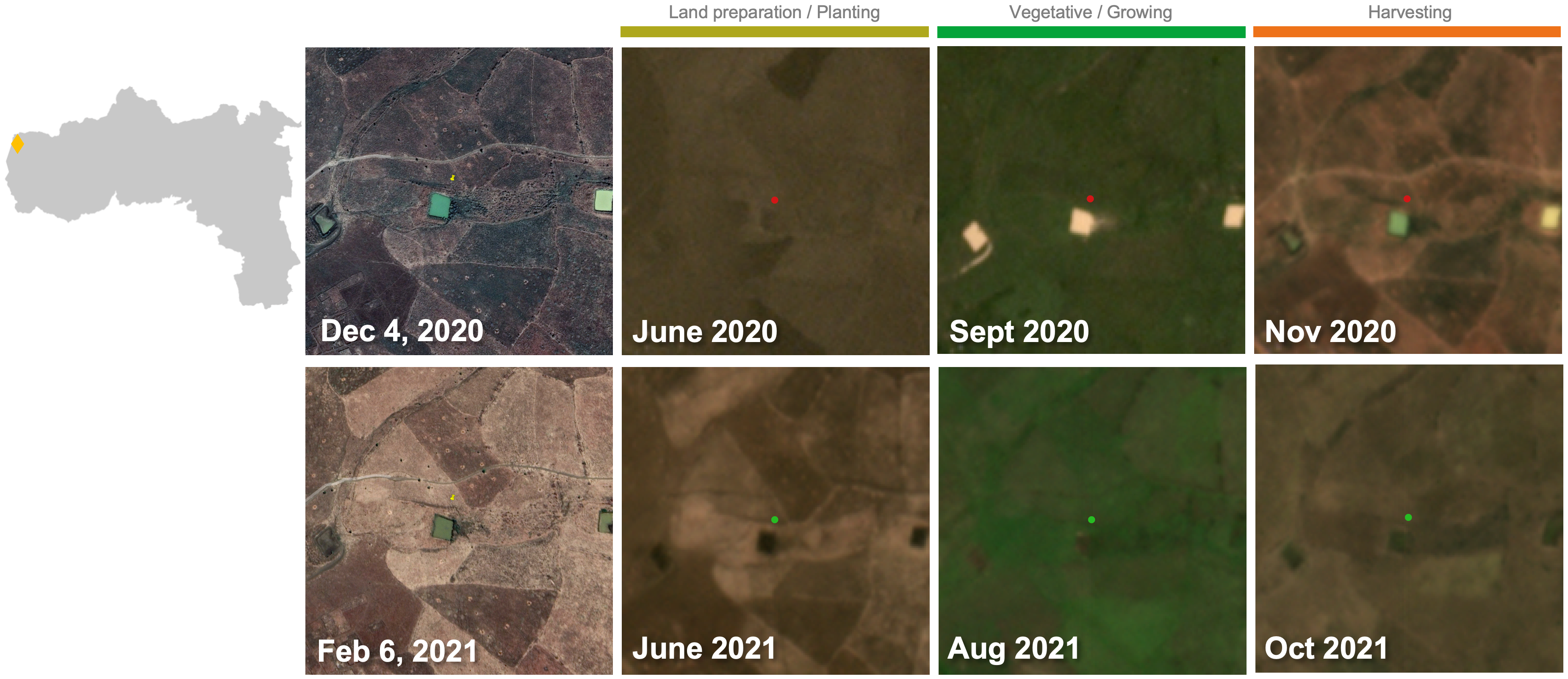}
    \caption{Example point labeled as planted cropland gain (id \#549). The reference point is in a field that appears fallow in 2020 but cultivated in 2021. In 2020, the field containing the point does not exhibit signs of land preparation, as it remains light brown in June 2020, or of vegetative growth, as it remains brownish/dull green in September 2020. In June 2021, the field and other fields surrounding it appear to have a darker brown color suggesting land preparation; in September, it appears bright green and uniformly vegetated; in November, it appears to have been cleared/harvested. The first column shows images from Airbus/CNES (2020) and Maxar (2021) from Google Earth while other columns show Planet Monthly mosaics from Collect Earth Online. The Google Earth images were taken outside of the growing season, but still provide helpful context showing the area in higher resolution than can be seen in the Planet mosaics.}
    \label{fig:p549-change-gain}
\end{figure}

\begin{figure}[ht]
    \centering
    \includegraphics[width=\textwidth]{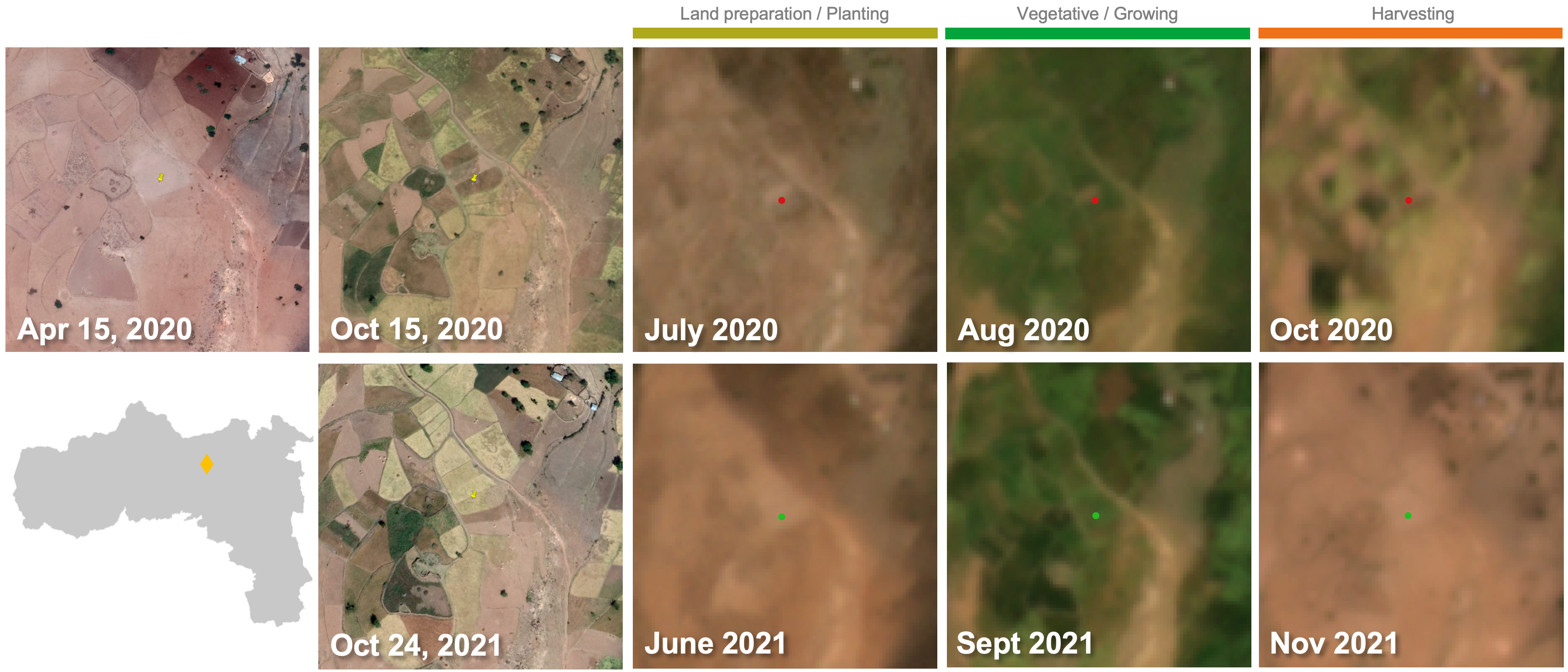}
    \caption{Example point labeled as planted cropland gain (id \#523). The reference point is in a field that appears fallow in 2020 but cultivated in 2021, which is clearly shown in the high-resolution Google Earth images from October 2020 versus 2021 shown in the second column. In August 2020, the field appears dull green compared to the brighter, uniform green shown in the September 2021 Planet mosaic. The first two columns show images from  Maxar from Google Earth while other columns show Planet Monthly mosaics from Collect Earth Online. }
    \label{fig:p523-change-gain}
\end{figure}

\begin{figure}[ht]
    \centering
    \includegraphics[width=\textwidth]{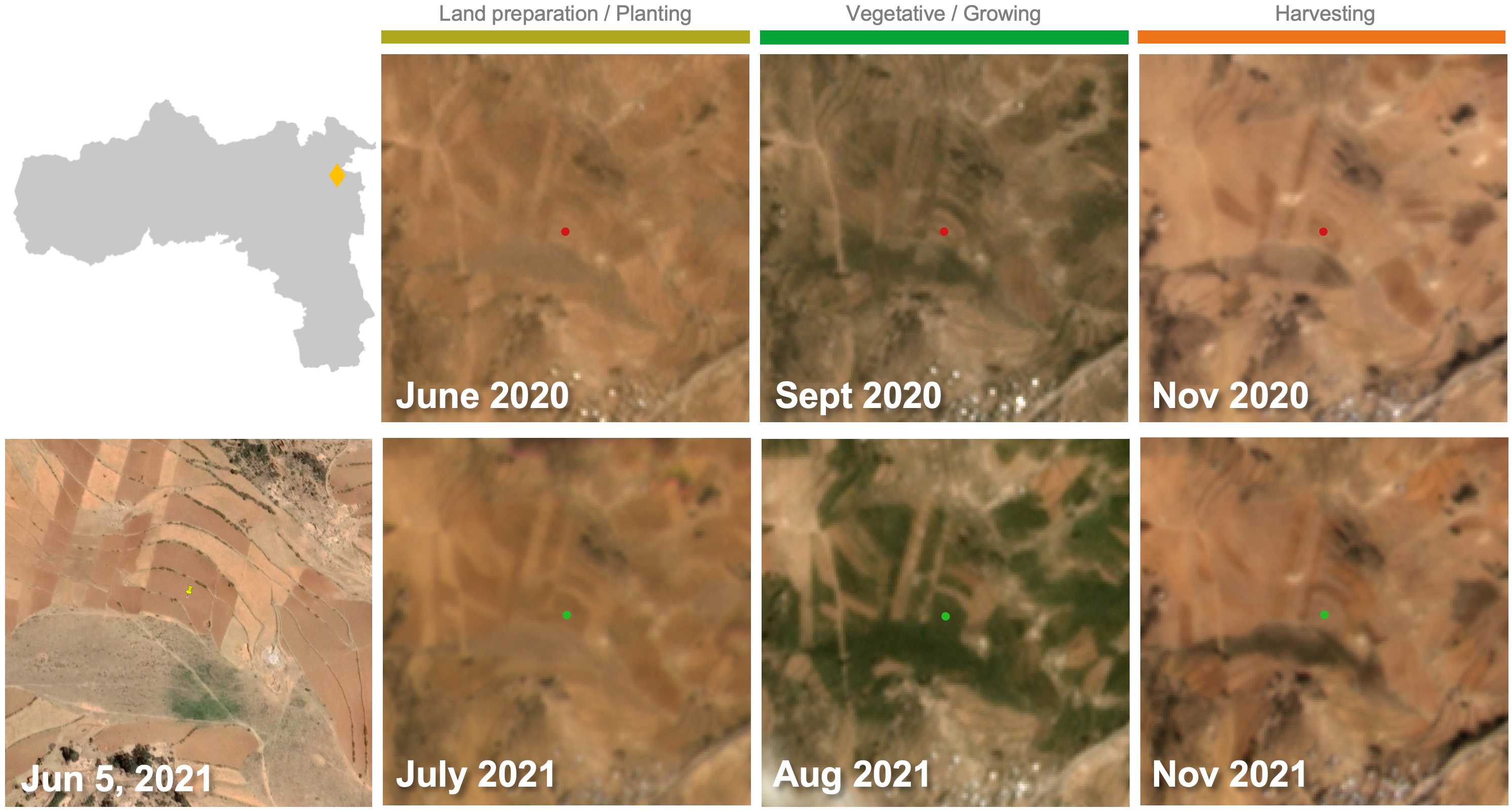}
    \caption{Example point labeled as planted cropland gain (id \#206). The reference point is in a field that appears fallow in 2020 but cultivated in 2021. The Google Earth image from June 2021 clearly shows that the field has been prepared for planting, which is also seen in the July 2021 Planet mosaic but not in 2020. The field appears densely green and vegetation in August 2021 but bare in the September 2020 mosaic. The first column shows an image from Maxar from Google Earth (no high-resolution image was available for 2020) while other columns show Planet Monthly mosaics from Collect Earth Online.}
    \label{fig:p206-change-gain}
\end{figure}

\begin{figure}[ht]
    \centering
    \includegraphics[width=\textwidth]{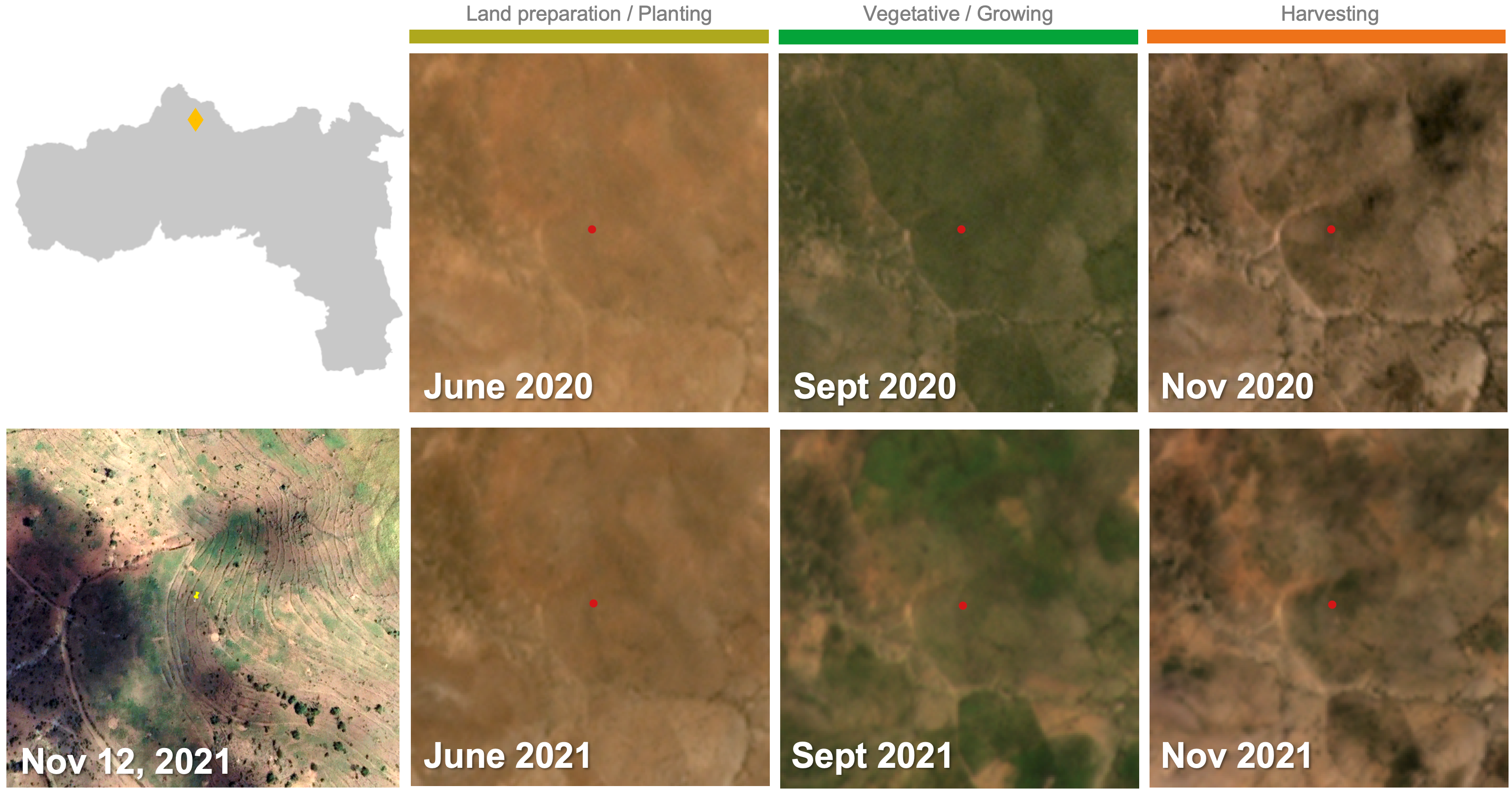}
    \caption{Example point labeled as stable non-crop (id \#0). The reference point is in a terraced area that could have been used for agriculture in prior years, but in 2020 and 2021 appears uncultivated. This can be seen in the high-resolution Google Earth image from November 2021 which shows patchy vegetation and the Planet Monthly mosaics which do not show signs of land preparation, dense vegetation during the growing period, or harvesting.}
    \label{fig:p0-stable-noncrop}
\end{figure}

\clearpage
\section{Sentinel-2 time series of cultivated and fallow fields}

\begin{figure}[ht]
    \centering
    \includegraphics[width=\textwidth]{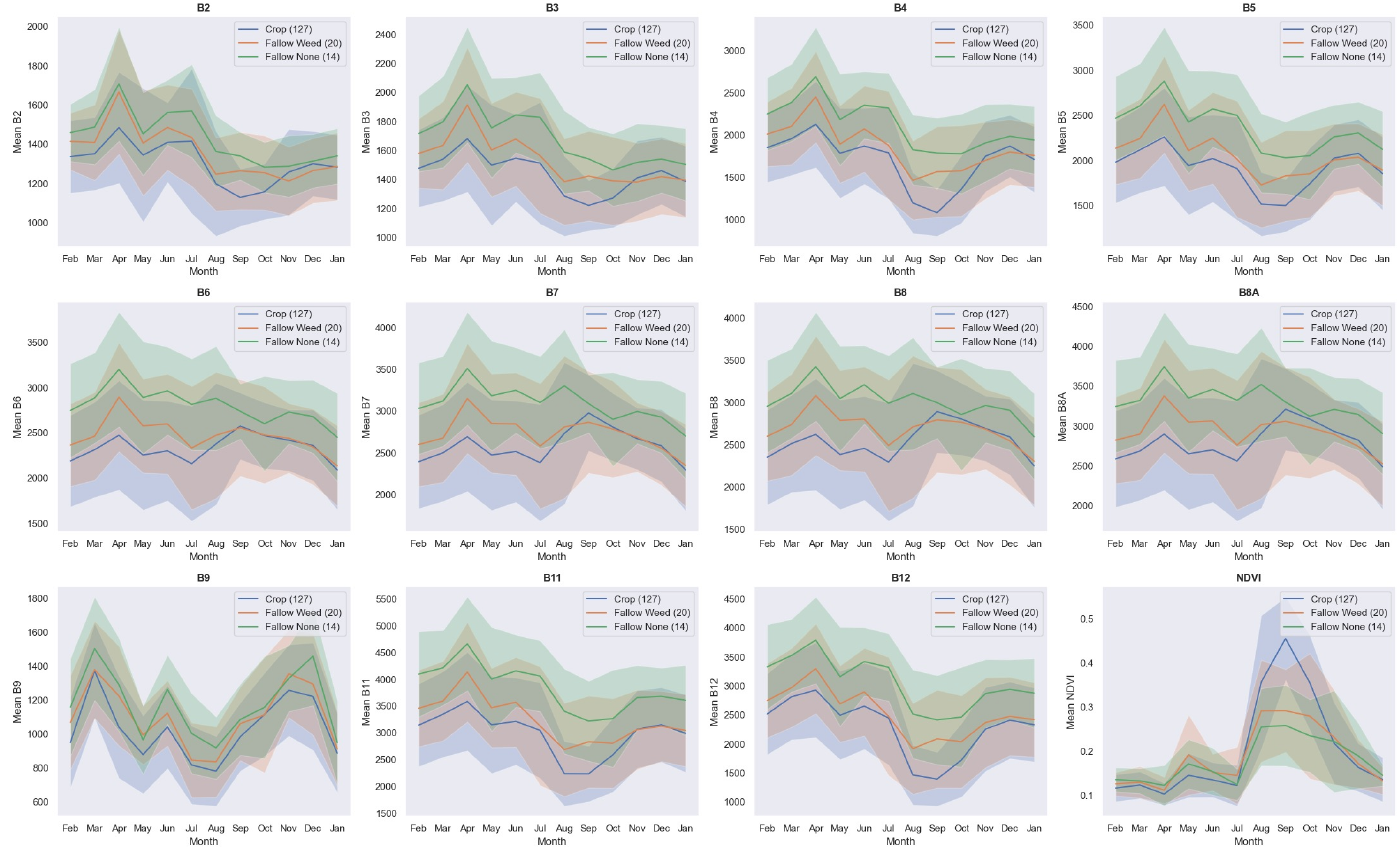}
    \caption{12-month time series in each Sentinel-2 input channel for ground-truth plots annotated as cultivated (``Crop'') and fallow with weeds (``Fallow Weed'') or fallow with no weeds/bare soil (``Fallow None''). Solid lines depict the mean and shaded regions depict the standard deviation of the time series for all samples in each band. }
    \label{fig:ghent-data-all}
\end{figure}

\clearpage
\section{Zone-scale confusion matrices}

\begin{figure}[ht]
     \centering
     \begin{subfigure}[b]{0.3\textwidth}
         \centering
         \includegraphics[width=\textwidth]{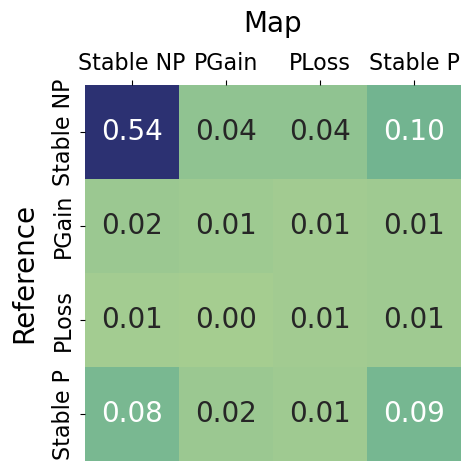}
         \caption{2020-2021 change map}
         \label{fig:cm-change-southern}
     \end{subfigure}
     \hfill
     \begin{subfigure}[b]{0.3\textwidth}
         \centering
         \includegraphics[width=\textwidth]{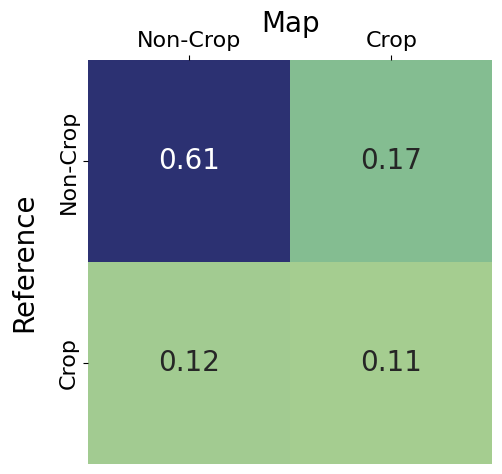}
         \caption{2020 map}
         \label{fig:cm-2020-southern}
     \end{subfigure}
    \hfill
     \begin{subfigure}[b]{0.3\textwidth}
         \centering
         \includegraphics[width=\textwidth]{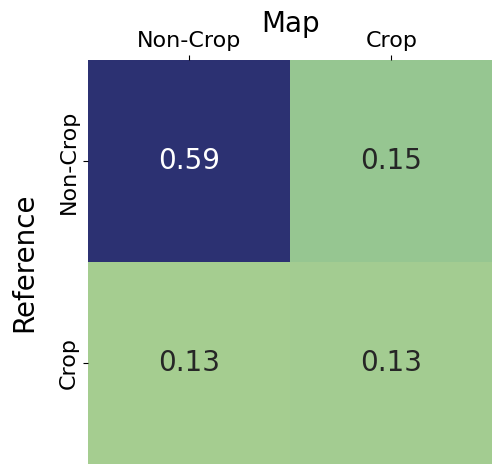}
         \caption{2021 map}
         \label{fig:cm-2021-southern}
     \end{subfigure}
        \caption{Confusion matrices for reference samples and change/annual cropland maps in Southern Tigray, expressed in terms of proportion of area as recommended in \cite{olofsson2014good}}
        \label{fig:confusion-southern}
\end{figure}

\begin{figure}[ht]
     \centering
     \begin{subfigure}[b]{0.3\textwidth}
         \centering
         \includegraphics[width=\textwidth]{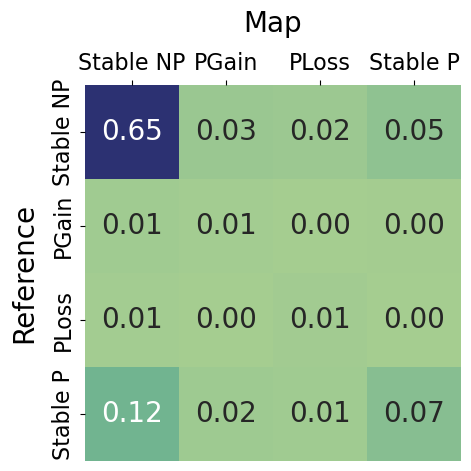}
         \caption{2020-2021 change map}
         \label{fig:cm-change-central}
     \end{subfigure}
     \hfill
     \begin{subfigure}[b]{0.3\textwidth}
         \centering
         \includegraphics[width=\textwidth]{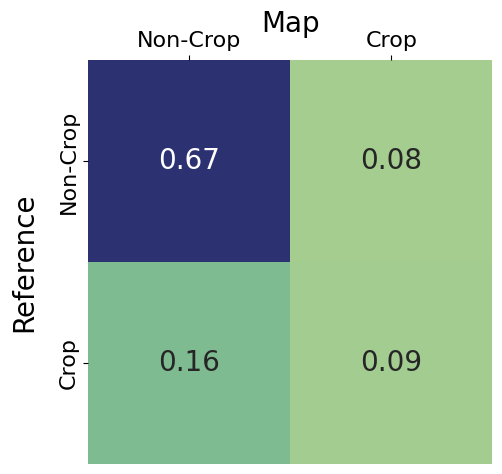}
         \caption{2020 map}
         \label{fig:cm-2020-central}
     \end{subfigure}
    \hfill
     \begin{subfigure}[b]{0.3\textwidth}
         \centering
         \includegraphics[width=\textwidth]{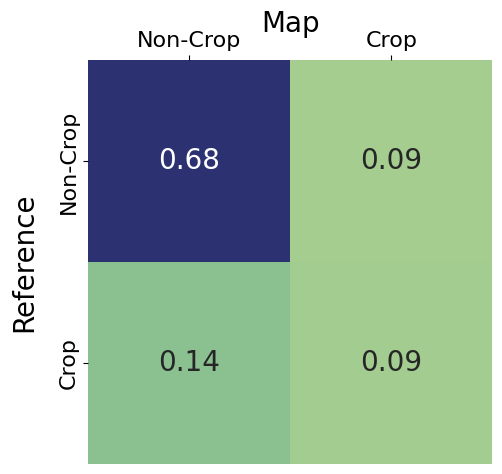}
         \caption{2021 map}
         \label{fig:cm-2021-central}
     \end{subfigure}
        \caption{Confusion matrices for reference samples and change/annual cropland maps in Central Tigray, expressed in terms of proportion of area as recommended in \cite{olofsson2014good}}
        \label{fig:confusion-central}
\end{figure}

\begin{figure}[ht]
     \centering
     \begin{subfigure}[b]{0.3\textwidth}
         \centering
         \includegraphics[width=\textwidth]{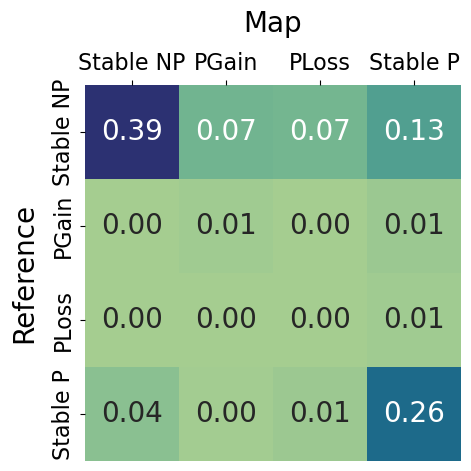}
         \caption{2020-2021 change map}
         \label{fig:cm-change-western}
     \end{subfigure}
     \hfill
     \begin{subfigure}[b]{0.3\textwidth}
         \centering
         \includegraphics[width=\textwidth]{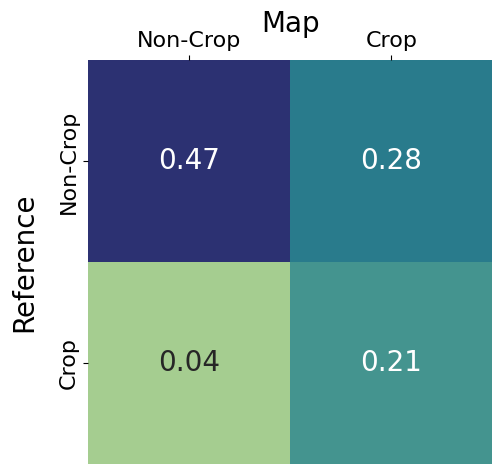}
         \caption{2020 map}
         \label{fig:cm-2020-western}
     \end{subfigure}
    \hfill
     \begin{subfigure}[b]{0.3\textwidth}
         \centering
         \includegraphics[width=\textwidth]{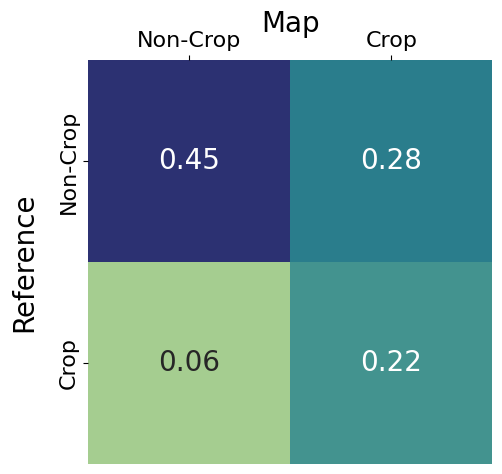}
         \caption{2021 map}
         \label{fig:cm-2021-western}
     \end{subfigure}
        \caption{Confusion matrices for reference samples and change/annual cropland maps in Western Tigray, expressed in terms of proportion of area as recommended in \cite{olofsson2014good}}
        \label{fig:confusion-western}
\end{figure}

\begin{figure}[ht]
     \centering
     \begin{subfigure}[b]{0.3\textwidth}
         \centering
         \includegraphics[width=\textwidth]{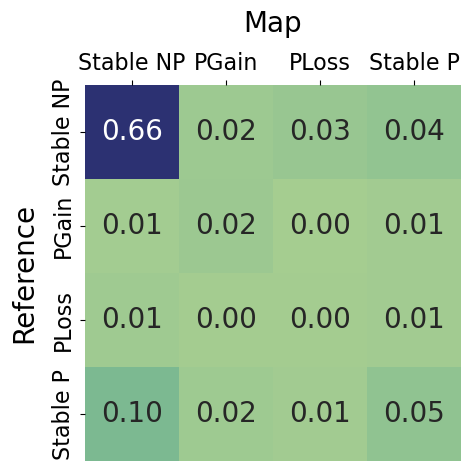}
         \caption{2020-2021 change map}
         \label{fig:cm-change-eastern}
     \end{subfigure}
     \hfill
     \begin{subfigure}[b]{0.3\textwidth}
         \centering
         \includegraphics[width=\textwidth]{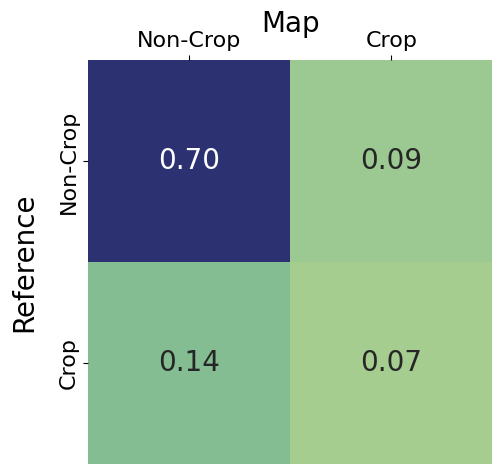}
         \caption{2020 map}
         \label{fig:cm-2020-eastern}
     \end{subfigure}
    \hfill
     \begin{subfigure}[b]{0.3\textwidth}
         \centering
         \includegraphics[width=\textwidth]{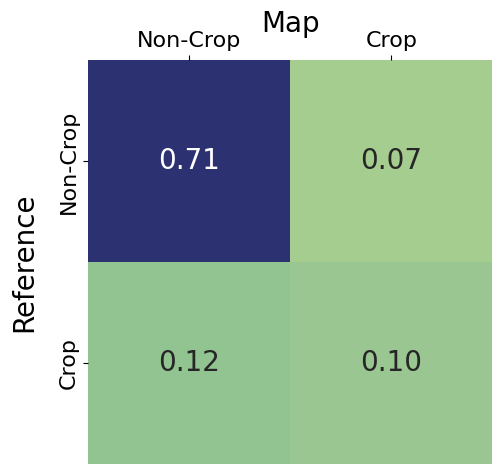}
         \caption{2021 map}
         \label{fig:cm-2021-eastern}
     \end{subfigure}
        \caption{Confusion matrices for reference samples and change/annual cropland maps in Eastern Tigray, expressed in terms of proportion of area as recommended in \cite{olofsson2014good}}
        \label{fig:confusion-eastern}
\end{figure}

\begin{figure}[ht]
     \centering
     \begin{subfigure}[b]{0.3\textwidth}
         \centering
         \includegraphics[width=\textwidth]{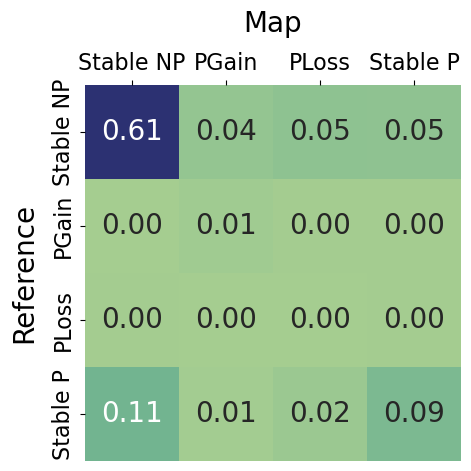}
         \caption{2020-2021 change map}
         \label{fig:cm-change-northwestern}
     \end{subfigure}
     \hfill
     \begin{subfigure}[b]{0.3\textwidth}
         \centering
         \includegraphics[width=\textwidth]{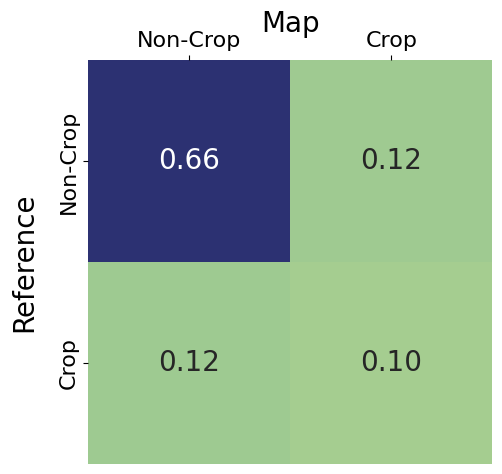}
         \caption{2020 map}
         \label{fig:cm-2020-northwestern}
     \end{subfigure}
    \hfill
     \begin{subfigure}[b]{0.3\textwidth}
         \centering
         \includegraphics[width=\textwidth]{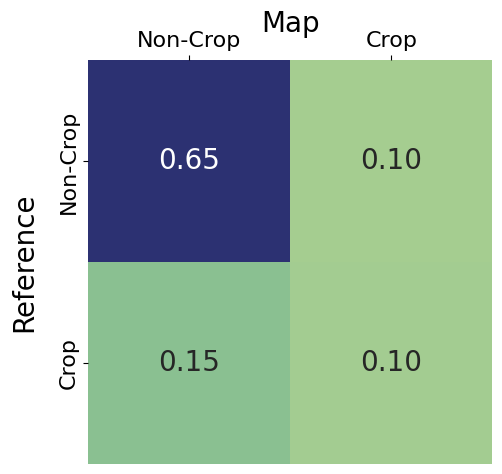}
         \caption{2021 map}
         \label{fig:cm-2021-northwestern}
     \end{subfigure}
        \caption{Confusion matrices for reference samples and change/annual cropland maps in North Western Tigray, expressed in terms of proportion of area as recommended in \cite{olofsson2014good}}
        \label{fig:confusion-northwestern}
\end{figure}

\clearpage
\section{Zone-scale change map accuracy}

\begin{table}[ht]
    \centering
\resizebox{\textwidth}{!}{%
\begin{tabular}{|c|c|c|c|c|}
        \hline
        Metric & Precision (UA) & Recall (PA) & True Positive Rate & False Positive Rate \\
        \hline \hline
        Stable cropland & $0.42 \pm 0.14$ & $0.44 \pm 0.10$ & $0.32$ & $0.11$\\
        Stable non-crop & $0.83 \pm 0.05$ & $0.75 \pm 0.04$ & $0.72$ & $0.38$  \\
        Cropland gain & $0.18 \pm 0.12$ & $0.25 \pm 0.16$ & $0.33$ & $0.10$  \\
        Cropland loss & $0.10 \pm 0.09$ & $0.27 \pm 0.25$ & $0.44$ & $0.11$ \\
        \hline
    \end{tabular}}
    \caption{Accuracy metrics for four-class cropland change reference sample in Southern zone of Tigray. Overall accuracy is $0.65 \pm 0.04$. Legend: UA = User's Accuracy, PA = Producer's Accuracy.} 
    \label{tab:change-accuracy-southern}
\end{table}

\begin{table}[ht]
    \centering
  \resizebox{\textwidth}{!}{%
  \begin{tabular}{|c|c|c|c|c|}
        \hline
        Metric & Precision (UA) & Recall (PA) & True Positive Rate & False Positive Rate \\
        \hline \hline
        Stable cropland & $0.56 \pm 0.19$ & $0.32 \pm 0.09$ & $0.22$ & $0.05$\\
        Stable non-crop & $0.82 \pm 0.05$ & $0.87 \pm 0.03$ & $0.81$ & $0.46$  \\
        Cropland gain & $0.13 \pm 0.11$ & $0.30 \pm 0.27$ & $0.56$ & $0.11$  \\
        Cropland loss & $0.22 \pm 0.17$ & $0.56 \pm 0.39$ & $0.71$ & $0.06$ \\
        \hline
    \end{tabular}}
    \caption{Accuracy metrics for four-class cropland change reference sample in Central zone of Tigray. Overall accuracy is $0.73 \pm 0.05$. Legend: UA = User's Accuracy, PA = Producer's Accuracy.} 
    \label{tab:change-accuracy-central}
\end{table}

\begin{table}[ht]
    \centering
   \resizebox{\textwidth}{!}{%
   \begin{tabular}{|c|c|c|c|c|}
        \hline
        Metric & Precision (UA) & Recall (PA) & True Positive Rate & False Positive Rate \\
        \hline \hline
        Stable cropland & $0.64 \pm 0.13$ & $0.83 \pm 0.08$ & $0.69$ & $0.12$\\
        Stable non-crop & $0.91 \pm 0.07$ & $0.59 \pm 0.05$ & $0.41$ & $0.11$  \\
        Cropland gain & $0.08 \pm 0.08$ & $0.29 \pm 0.36$ & $0.60$ & $0.18$  \\
        Cropland loss & $0.00 \pm 0.00$ & $0.00 \pm 0.00$ & $0.00$ & $0.23$ \\
        \hline
    \end{tabular}}
    \caption{Accuracy metrics for four-class cropland change reference sample in Western zone of Tigray. Overall accuracy is $0.66 \pm 0.06$. Legend: UA = User's Accuracy, PA = Producer's Accuracy.} %
    \label{tab:change-accuracy-western}
\end{table}

\begin{table}[ht]
    \centering
  \resizebox{\textwidth}{!}{%
  \begin{tabular}{|c|c|c|c|c|}
        \hline
        Metric & Precision (UA) & Recall (PA) & True Positive Rate & False Positive Rate \\
        \hline \hline
        Stable cropland & $0.43 \pm 0.21$ & $0.28 \pm 0.13$ & $0.30$ & $0.09$\\
        Stable non-crop & $0.85 \pm 0.07$ & $0.87 \pm 0.03$ & $0.79$ & $0.34$  \\
        Cropland gain & $0.36 \pm 0.21$ & $0.57 \pm 0.31$ & $0.73$ & $0.09$  \\
        Cropland loss & $0.07 \pm 0.14$ & $0.11 \pm 0.21$ & $0.17$ & $0.08$ \\
        \hline
    \end{tabular}}
    \caption{Accuracy metrics for four-class cropland change reference sample in Eastern zone of Tigray. Overall accuracy is $0.74 \pm 0.06$. Legend: UA = User's Accuracy, PA = Producer's Accuracy.} 
    \label{tab:change-accuracy-eastern}
\end{table}

\begin{table}[ht]
    \centering
\resizebox{\textwidth}{!}{%
\begin{tabular}{|c|c|c|c|c|}
        \hline
        Metric & Precision (UA) & Recall (PA) & True Positive Rate & False Positive Rate \\
        \hline \hline
        Stable cropland & $0.61 \pm 0.15$ & $0.39 \pm 0.09$ & $0.39$ & $0.08$\\
        Stable non-crop & $0.84 \pm 0.06$ & $0.82 \pm 0.03$ & $0.69$ & $0.34$  \\
        Cropland gain & $0.19 \pm 0.15$ & $0.61 \pm 0.34$ & $0.62$ & $0.08$  \\
        Cropland loss & $0.02 \pm 0.04$ & $0.18 \pm 0.35$ & $0.33$ & $0.16$ \\
        \hline
    \end{tabular}}
    \caption{Accuracy metrics for four-class cropland change reference sample in North Western zone of Tigray. Overall accuracy is $0.71 \pm 0.05$. Legend: UA = User's Accuracy, PA = Producer's Accuracy.} 
    \label{tab:change-accuracy-northwestern}
\end{table}

\clearpage
\section{Conflict buffer confusion matrices}

\begin{figure}[ht]
     \centering
     \begin{subfigure}[b]{0.3\textwidth}
         \centering
         \includegraphics[width=\textwidth]{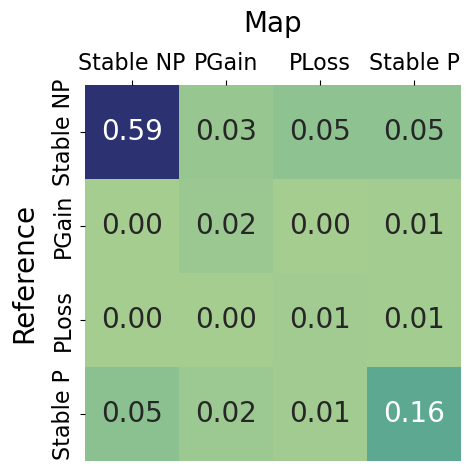}
         \caption{Inside conflict buffer}
         \label{fig:cm-inside-buffer}
     \end{subfigure}
     \begin{subfigure}[b]{0.3\textwidth}
         \centering
         \includegraphics[width=\textwidth]{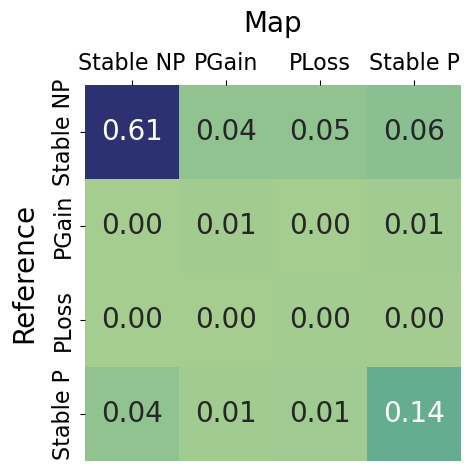}
         \caption{Outside conflict buffer}
         \label{fig:cm-outside-buffer}
     \end{subfigure}
        \caption{Confusion matrices for reference samples and change maps inside and outside 5km conflict event buffer, expressed in terms of proportion of area as recommended in \cite{olofsson2014good}}
        \label{fig:confusion-buffer}
\end{figure}

\clearpage
\section{Conflict buffer change map accuracy}

\begin{table}[ht]
    \centering
 \resizebox{\textwidth}{!}{%
 \begin{tabular}{|c|c|c|c|c|}
        \hline
        Metric & Precision (UA) & Recall (PA) & True Positive Rate & False Positive Rate \\
        \hline \hline
        Stable cropland & $0.74 \pm 0.13$ & $0.66 \pm 0.11$ & $0.55$ & $0.07$\\
        Stable non-crop & $0.92 \pm 0.06$ & $0.83 \pm 0.04$ & $0.63$ & $0.10$  \\
        Cropland gain & $0.30 \pm 0.14$ & $0.75 \pm 0.30$ & $0.87$ & $0.15$  \\
        Cropland loss & $0.11 \pm 0.10$ & $0.52 \pm 0.45$ & $0.67$ & $0.15$ \\
        \hline
    \end{tabular}}
    \caption{Accuracy metrics for four-class cropland change reference sample inside 5km buffer of conflict events in Tigray. Overall accuracy is $0.78 \pm 0.05$. Legend: UA = User's Accuracy, PA = Producer's Accuracy.} 
    \label{tab:change-accuracy-buffer5km}
\end{table}

\begin{table}[ht]
    \centering
    \resizebox{\textwidth}{!}{%
    \begin{tabular}{|c|c|c|c|c|}
        \hline
        Metric & Precision (UA) & Recall (PA) & True Positive Rate & False Positive Rate \\
        \hline \hline
        Stable cropland & $0.66 \pm 0.10$ & $0.68 \pm 0.07$ & $0.51$ & $0.06$\\
        Stable non-crop & $0.93 \pm 0.03$ & $0.80 \pm 0.02$ & $0.65$ & $0.15$  \\
        Cropland gain & $0.13 \pm 0.07$ & $0.43 \pm 0.25$ & $0.67$ & $0.14$  \\
        Cropland loss & $0.07 \pm 0.05$ & $0.66 \pm 0.47$ & $0.88$ & $0.15$ \\
        \hline
    \end{tabular}}
    \caption{Accuracy metrics for four-class cropland change reference sample outside 5km buffer of conflict events in Tigray. Overall accuracy is $0.77 \pm 0.03$. Legend: UA = User's Accuracy, PA = Producer's Accuracy.} 
    \label{tab:change-accuracy-outsidebuffer}
\end{table}

\clearpage
\section{Effect of sample size on conflict buffer estimates}
Since the reference sample size for the area inside the conflict buffer ($n=219$) was smaller than the sample size for the area outside the conflict buffer ($n=582$), it is possible that the higher upper bound of loss in the conflict buffer reported in Table \ref{tab:change-area-buffer} is due to a relatively smaller sample size. To test this, we performed an experiment to estimate the cropland loss area outside the conflict buffer with a smaller sample size equal to that of the conflict buffer. We randomly sub-sampled 219 points from the non-conflict 582 reference samples, then calculated the cropland loss area and CI using this sample. We repeated this 10 times using different random seeds, since the result may vary depending on the samples drawn. We calculated the median and mean cropland loss area and CI over all random seeds. We report the results in Table \ref{tab:conflict-buffer-seeds}.

\begin{table}[ht]
\centering
\begin{tabular}{|c|c|c|c|c|}
\hline
Seed & Cropland loss area (kha) & OA & UA & PA \\
\hline \hline
1 & $26 \pm 28$ & 0.74 $\pm$ 0.05 & 0.10 $\pm$ 0.11 & 1.00 $\pm$ 0.00 \\
10 & $54 \pm 61$ & 0.79 $\pm$ 0.05 & 0.10 $\pm$ 0.09 & 0.47 $\pm$ 0.54 \\
100 & $22 \pm 43$ & 0.77 $\pm$ 0.05 & 0.00 $\pm$ 0.00 & 0.00 $\pm$ 0.00 \\
1000 & $21 \pm 23$ & 0.76 $\pm$ 0.05 & 0.08 $\pm$ 0.09 & 1.00 $\pm$ 0.00 \\
10000 & $41 \pm 48$ & 0.77 $\pm$ 0.05 & 0.08 $\pm$ 0.08 & 0.47 $\pm$ 0.56 \\
100000 & $33 \pm 27$ & 0.77 $\pm$ 0.05 & 0.13 $\pm$ 0.11 & 1.00 $\pm$ 0.00 \\
2 & $15 \pm 20$ & 0.75 $\pm$ 0.05 & 0.06 $\pm$ 0.08 & 1.00 $\pm$ 0.00 \\
20 & $28 \pm 27$ & 0.76 $\pm$ 0.05 & 0.11 $\pm$ 0.10 & 1.00 $\pm$ 0.00 \\
200 & $51 \pm 70$ & 0.73 $\pm$ 0.05 & 0.07 $\pm$ 0.07 & 0.33 $\pm$ 0.50 \\
2000 & $29 \pm 27$ & 0.78 $\pm$ 0.05 & 0.11 $\pm$ 0.11 & 1.00 $\pm$ 0.00 \\
\hline
Median & $29 \pm 28$ & -- & -- & -- \\
Mean & $32 \pm 37$ & -- & -- & -- \\
\hline
\end{tabular}
\caption{Cropland loss area, overall accuracy (OA), user's accuracy (UA), and producer's accuracy (PA) for random subsets of size $n=219$ from reference sample in region outside of the 5 km conflict buffer.}
\label{tab:conflict-buffer-seeds}
\end{table}

\begin{table}[ht]
\centering
\begin{tabular}{|l|c|c|}
\hline
 & Cropland loss area (kha) & Percentage of total area \\
\hline \hline
Inside buffer (n=219) & 19 $\pm$ 17 & 0--3\% \\
Outside buffer (n=582) & 28 $\pm$ 23 & 0--1\% \\
Outside buffer (n=219, median) & 29 $\pm$ 28 & 0--1\% \\
Outside buffer (n=219, mean) & 32 $\pm$ 37 & 0--2\% \\
\hline
\end{tabular}
\caption{Comparison of cropland loss area estimates inside and outside the 5 km conflict buffer for full sample sizes (reported in Table \ref{tab:change-area-buffer}) and equal sample sizes.}
\label{tab:conflict-buffer-sample-test}
\end{table}

\end{document}